\documentclass[superscriptaddress,aps,pra,twocolumn,showpacs,nofootinbib,longbibliography,notitlepage]{revtex4-2}
\usepackage{etex}
\usepackage{amsmath,amssymb,amsthm}
\usepackage[colorlinks=true,citecolor=blue,urlcolor=blue]{hyperref}
\usepackage[pdftex]{graphicx}
\usepackage{times,txfonts}
\usepackage{braket}
\usepackage{color}
\usepackage{natbib}
\usepackage{amsmath,blkarray}
\usepackage{mathtools}
\usepackage{latexsym}
\usepackage{tabularx, booktabs}
\usepackage{graphics,epstopdf}
\usepackage{graphicx}
\usepackage{float}
\usepackage{graphicx}
\usepackage{amsfonts}
\usepackage{subcaption}
\usepackage{color,soul}

\newcommand{\be}{\begin{equation}}
	\newcommand{\ee}{\end{equation}}
\newcommand{\ba}{\begin{eqnarray}}
	\newcommand{\ea}{\end{eqnarray}}

\begin{document}
	\title{Device-independent self-testing of unsharp measurements}
	\author{ Prabuddha Roy }
	\email{prabuddharoy.94@gmail.com}
		\affiliation{National Institute of Technology Patna, Ashok Rajpath, Patna, Bihar 800005, India}
	\author{ A. K. Pan }
	\email{akp@phy.iith.ac.in}
	\affiliation{National Institute of Technology Patna, Ashok Rajpath, Patna, Bihar 800005, India}
 \affiliation{Department of Physics, Indian Institute of Technology Hyderabad, Telengana-502284, India }
	
	\begin{abstract}
Semi-device-independent certification of an unsharp instrument has recently been demonstrated [\href{https://iopscience.iop.org/article/10.1088/1367-2630/ab3773}{New J. Phys. {\bf 21}, 083034 (2019)}] based on the sequential sharing of quantum advantages in a prepare-measure communication game by assuming the system to be qubit. In this work, we provide device-independent (DI) self-testing of the unsharp instrument through the quantum violation of two Bell inequalities where the devices are uncharacterized and the dimension of the system remains unspecified.   We introduce an elegant sum-of-squares approach to derive the dimension-independent optimal quantum violation of  Bell inequalities which plays a crucial role. Note that the standard Bell test cannot self-test  the post-measurement states and consequently cannot self-test unsharp instrument. The sequential Bell test possess the potential to self-test an unsharp instrument. We demonstrate that there exists a trade-off between the maximum sequential quantum violations of the Clauser-Horne-Shimony-Holt inequality, and they form an optimal pair that enables the DI self-testing of the entangled state, the observables, and the unsharpness parameter. Further, we extend our study to the case of elegant Bell inequality and we argue that it has two classical bounds - the local bound and the non-trivial preparation non-contextual bound, lower than the local bound. Based on the sharing of preparation contextuality by three independent sequential observers, we demonstrate the DI self-testing of two unsharpness parameters. Since an actual experimental scenario involves losses and imperfection, we demonstrate robustness of our certification to noise.
\end{abstract}
\maketitle
\section{Introduction}
Measurement plays a pivotal role in quantum theory which is in stark contrast to classical theory. The textbook version of a quantum measurement is modeled by a set of orthogonal projectors belonging to the Hilbert space. However, there exist more general measurements defined in terms of positive-operator-valued measures (POVMs) satisfying the completeness relation. Note that the POVMs can be defined in many different ways. However, in this paper, we are concerned about those POVMs which are noisy or unsharp variants of the projective measurements. 

In a sharp projective measurement \cite{von}, the system collapses to one of the eigenstates of the measured observable, and the system is completely disturbed by the process of sharp measurement, so that, no residual coherence remains in the system. On the other hand, in the case of unsharp measurements that are characterized by the POVMs, the system is less disturbed compared to the sharp projective measurement. Since a projective measurement maximally disturbs the quantum system and hence extracts more information from the system compared to POVMs, one may think that a sharp measurement is more advantageous in information processing tasks. However, there exist certain information processing tasks where POVMs are proven to be more useful compared to sharp measurement.

Advantage of POVMs over sharp measurement has been explored in the context of quantum state discrimination \cite{std1,std2}, randomness certification \cite{acin,pan21,and18,ran1}, quantum tomography\cite{derka}, state estimation \cite{bergou}, quantum cryptography \cite{renes}, information acquisition from a quantum source \cite{jozsa}, quantum violation of certain Bell inequalities \cite{vertesi}, and many more. There is one more advantage that is particularly relevant in the present work is the sequential sharing of various forms of quantum correlations \cite{silva2015,sasmal2018,bera2018,kumari2019,brown2020,Zhang21}. For example, a sequential Bell test by multiple independent observers inevitably requires the prior observers to perform the unsharp quantum measurements.

In this work, we aim to provide device-independent (DI) self-testing of the unsharp measurements, which are noisy variants of the sharp projective measurements. Self-testing \cite{self1,mckague16,mckague17,supic20,supic21} is the strongest form of certification protocol where the devices are treated as black boxes. Also, the dimension of the quantum system is assumed to be finite but unbounded. In that, observed experimental statistics uniquely certify the state and measurement observables of an unknown dimension. DI self-testing is advantageous over the standard certification methods such as those based on quantum tomography. Essentially, a self-testing protocol requires optimal quantum violation of a suitable Bell's inequality \cite{bell}. For example, optimal violation of Clauser-Horne-Shimony-Holt (CHSH) inequality self-tests the maximally entangled state and mutually anticommuting local observables in any arbitrary dimension. Note that the DI certification encounters practical challenges arising from the requirement of a loophole-free Bell test. Such tests have recently been realized \cite{lf1,lf2,lf3,lf4,lf5} enabling experimental demonstrations of DI certification of randomness \cite{Yang, Peter}. Of late, the DI certification is used as a resource for secure quantum key distribution \cite{bar05,acin06,acin07,pir09}, randomness certification \cite{pir10,nieto,col12}, witnessing Hilbert space dimension \cite{wehner,gallego,ahrens,brunnerprl13,sik16prl,cong17,pan2020} and for achieving advantages in communication complexity tasks \cite{complx1}.

We provide DI self-testing schemes to certify the unsharpness parameters through the quantum violation of two well-known Bell inequalities viz., the CHSH inequality \cite{bell} and the elegant Bell inequality \cite{gisin}. Note that optimal quantum violations of such  Bell inequalities can be obtained only for sharp measurement. Any value less than the optimal quantum value may originate due to various reasons. It may come from the unsharp measurements of local observables but may also come from the nonideal preparation of the state or the inappropriate choices of the local observables than the ones required for the optimal quantum violation. However, a more serious issue regarding DI self-testing of unsharp measurement through a Bell test arises due to Naimark's theorem. It states that every non-projective measurement can be modeled as a projective measurement in a larger Hilbert space by introducing suitable ancilla in a higher dimension. Since in the DI Bell test, there is no bound on the dimension, a stubborn physicist may argue that the sub-optimal quantum violation is arising due to the inappropriate choice of observables in a higher dimension and \emph{not} due to the unsharpness of the measurement. Hence, to self-test unsharp measurements, one needs to introduce a protocol that certifies the state, the observables, and the unsharpness parameters without referring to the dimension of the system.

Against this backdrop, we demonstrate that the sequential Bell test has such potential where the sub-optimal sequential quantum violations of a Bell inequality by multiple independent observers on one side can enable such a self-testing. Such a scheme successfully bypasses the constraints that would arise from Naimark's theorem as the optimization of sequential Bell expression is performed without imposing any constraint on the dimension of Hilbert space. As far as our knowledge goes, the DI self-testing of non-projective measurements through the Bell test has not hitherto been demonstrated. However,  semi-DI certification of non-projective measurements has been demonstrated by using the qubit system. In \cite{gomez16,mir2019,samina2020}, the extremal qubit POVMs were experimentally certified based on the theoretical proposal \cite{acin}. Those experiments \cite{gomez16,mir2019,samina2020,tava20exp} do certify the non-projective character of measurement, but not how it relates to a specific target POVM. The semi-DI certification of qubit unsharp measurements (noisy variants of projective measurement) in the prepare-measure scenario has recently been proposed \cite{mohan,mercin,wei,sumit}. The proposal of \cite{mohan} has been  experimentally verified \cite{tava20exp,anwar2020,fole2020}. 

Specifically, in our self-testing protocol, two sequential sub-optimal quantum violations of a Bell inequality form an optimal pair powering the DI certifications of the state, the local observables, and the unsharpness parameter of one of the two parties. We also note here that all the previous works that demonstrated the sharing of various quantum correlations \cite{brown2020,Zhang2021,kumari2019,Debarshi,Shashank,Akshata,saptarshi,sumit,Cheng2021,Mao2022,Cheng22}, the dimension of the system was assumed. In contrast, throughout this work, we impose no bound on the dimension of the Hilbert space, and we consider that the measurement devices are black boxes. We first demonstrate that, at most, two independent sequential observers on one side can violate the CHSH inequality.   We simultaneously maximize the quantum violations of CHSH inequality for two sequential observers on one side and demonstrate that there is a trade-off between the two sub-optimal quantum violations. This, in turn, certifies the state, and the observables for both the sequential observers and  the unsharpness parameter of the first observer. Since in a practical implementation there remains imperfection, we show how a range of an unsharpness parameter can be self-tested in that scenario. 

The protocol developed for the  CHSH inequality is further extended to the case of elegant Bell inequality,  where we demonstrate that, at most, two observers can share the quantum advantage when considering the inequality's local bound. However, we argue that the elegant Bell inequality has two classical bounds, the local and the preparation non-contextual bound, and the latter is smaller than the former. We show that if we consider the preparation non-contextual bound of the elegant Bell inequality, then at most three observers can share the quantum advantage. Further, we demonstrate that if the quantum advantage is extended to a third sequential observer, then the range of the values of the unsharpness parameter for the first observer can be more restricted, thereby self-testing a narrow range of the values of the unsharpness parameter in the sub-optimal scenario.

This paper is organized as follows. In Sec. II, we demonstrate the optimal quantum violation of CHSH inequality using an elegant sum-of-squares approach. In section III, we explicitly show the sequential violation of CHSH inequality and self-testing of the unsharpness parameter. In Sec. IV, we briefly introduce the notion of the preparation non-contextuality in an ontological model and provide the preparation non-contextual bound of elegant Bell inequality. In Sec. V, we extend the sharing of preparation contextuality by three sequential observers based on the quantum violation of elegant Bell inequality and demonstrate the self-testing of two unsharpness parameters along with the states and the observables. Finally,  we discuss our results in Sec. VI.

\section{Optimal quantum violation of CHSH inequality}
	The CHSH scenario consists of two space-like separated parties (say, Alice and Bob) who share a common physical system. Alice (Bob) performs local measurements on her (his) subsystem upon receiving inputs  $x\in \{1,2\} (y \in \{1,2\})$, and returns outputs $a \in \{0,1\}(b \in \{0,1\})$. Representing $M_{x}$ and $N_{y}$ as respective dichotomic observables of Alice and Bob, the CHSH form of Bell's inequality can be written as
	\begin{equation}
		\label{bell}
		\mathcal{B}=\left(M_{1}+M_{2}\right) N_{1} +\left(M_{1}-M_{2}\right) N_{2}\leq 2.
	\end{equation}
	The optimal quantum value of the CHSH expression is $\left(\mathcal{B}\right)^{opt}_{Q}=2\sqrt{2}$, commonly known as Tirelson bound \cite{cirelson}. The optimal value can be obtained when the shared state is maximally entangled in a two-qubit system and the local qubit observables are mutually anticommuting. However, one does not need to impose the bound on the dimension to obtain the optimal value. Also, the measurement of Alice and Bob remains uncharacterized. In other words, the optimal value $\left(\mathcal{B}\right)^{opt}_Q$ can be achieved if Alice and Bob perform dichotomic measurements on a maximally entangled state $\ket{\psi}_{AB} \in \mathcal{C}^{d}\otimes \ \mathcal{C}^{d}$ where $d\geq 2$ is arbitrary. 
 
 For our purpose, we provide a derivation of the $\left(\mathcal{B}\right)^{opt}_{Q}$ without imposing a bound on the Hilbert space dimension by introducing an elegant sum-of-squares (SOS) approach. Let us assume that $(\mathcal{B})_{Q}\leq \Omega_{2}$, where  $\Omega_{2}$ is a positive quantity and the upper bound of $(\mathcal{B})_{Q}$. Equivalently, one can argue that there is a positive semi-definite operator $\eta \geq 0$, that can be expressed as 
  \ba
 \langle \eta\rangle_{Q}=\Omega_{2} -(\mathcal{B})_{Q} 
 \ea
 for a quantum state $|\psi\rangle_{AB}$. This can be proven by considering two suitable positive operators, $L_{1}$ and $L_{2}$, which are polynomial functions of   $M_{x}$ and $N_y$, so that 

\begin{align}
	\label {g1}
	\eta =\frac{1}{2} \left( \omega_{1} L_1^\dagger L_1+\omega_{2} L_2^\dagger L_2\right). 
\end{align}
For our purpose, we suitably choose $ L_1 $ and $L_{2}$ as  

\begin{align}
	\label{mi}
	L_{1}|\psi\rangle_{AB}=\frac{M_{1}+M_{2}}{\omega_{1}} |\psi\rangle_{AB} -N_{1} |\psi\rangle_{AB}\\
	\nonumber
	L_{2}|\psi\rangle_{AB}=\frac{M_{1}-M_{2}}{\omega_{1}} |\psi\rangle_{AB} -N_{2} |\psi\rangle_{AB}
\end{align}
where $\omega_{1} =||\left(M_{1}+M_{2}\right)|\psi\rangle_{AB}||_{2}$ an $\omega_{2} =||\left(M_{1}-M_{2}\right)|\psi\rangle_{AB}||_{2}$. Here $||.||_{2}$ is the Frobenius norm of a vector, $||\ \mathcal{O}\ ||_2=\sqrt{Tr[\mathcal{O}^{\dagger}\mathcal{O} \rho]}$. Plugging Eq. (\ref{mi}) into Eq. (\ref{g1}) and by noting that $M_{x}^{\dagger} M_{x}=N_y^{\dagger} N_y=\mathbb{I} $, we get

\begin{align}
	(\mathcal{B})_{Q}= \left(\omega_{1}+\omega_{2}\right) -\langle \eta\rangle_{Q} .
\end{align}
Optimal value of $(\mathcal{B})_{Q}$ can be obtained when $\langle\eta\rangle_{Q}= 0$, i.e.,   

\begin{eqnarray}
	\label{optbnn}
	(\mathcal{B})_{Q}^{opt} &=&\underset{{}}{max}\left(\omega_{1}+\omega_{2}\right)\\
	\nonumber
	&=&\underset{{}}{max}\left(\sqrt{2+\langle \{M_1, M_2\}\rangle}+\sqrt{2-\langle \{M_1, M_2\}\rangle}\right) .
\end{eqnarray}
Thus, the maximization requires $ \{M_1, M_2\}=0$ implying Alice's observables have to be anticommuting. In turn, we find the values $\omega_{1}=\omega_{2}=\sqrt{2}$, and consequently the optimal value $(\mathcal{B})_{Q}^{opt}=2\sqrt{2}$. The explicit conditions for the optimization are $L_{1}|\psi\rangle_{AB}=0$ and $L_{2}|\psi\rangle_{AB}=0$, i.e., $N_{1}=(M_{1}+M_{2})/\sqrt{2}$ and $N_{2}=(M_{1}-M_{2})/\sqrt{2}$. It can be easily checked that $\{N_1, N_2\}=0$, i.e., Bob's observables are also anticommuting. 

    Note also that for the state $\rho_{AB}=\ket{\psi_{AB}}\bra{\psi_{AB}} \in \mathcal{C}^d \otimes \mathcal{C}^d $, optimal violation requires $Tr[N_{1}\otimes N_{1}\rho_{AB}]=Tr[N_{2}\otimes N_{2}\rho_{AB}]=1$, which again confirms that $\rho_{AB}$ has to be a pure state. Let us choose the state in Hilbert-Schmidt form as
\ba
\label{mentngled}
\rho_{AB}=\frac{1}{d^{2}}\Big[\mathbb{I}\otimes\mathbb{I}+\sum_{i=1}^{d^2-1}N_{i}\otimes N_{i}\Big]
\ea
where $\{N_{i},N_{j}\}=0$ and consequently $[N_{i}\otimes N_{i},N_{j}\otimes N_{j}]=0$ for any arbitrary dimension $d$. For a density matrix $\rho_{AB}$, $Tr[\rho_{AB}]=1$  has to be satisfied. This in turn provides $Tr[N_{1}]=Tr[N_{2}]=0$. Also, $Tr[\rho_{AB}^{2}]=1$ ensures that the observables in the summation in Eq. (\ref{optbnn}) contains full set of mutually anticommuting observables $\{N_{i}\otimes N_{i}\}$. Consequently,  $Tr_{A}[\rho_{AB}]=Tr_{B}[\rho_{AB}]=\frac{\mathbb{I}}{d}$, i.e., partial trace of $\rho_{AB}$ is maximally mixed state for both Alice and Bob. Thus, the optimal violation of the CHSH inequality is achieved for the maximally entangled state $\ket{\psi_{AB}}$. We thus derive the optimal quantum value $(\mathcal{B})_{Q}^{opt}$ which uniquely certifies the state and observables. The entire derivation is done without assuming the dimension of the system.

	\section{Sequential sharing of nonlocality and self-testing of unsharpness parameter}	

The sequential Bell-CHSH test in the DI scenario is depicted in \figurename{ 1}. There is only one Alice, who always performs sharp measurement and arbitrary $k$ number of sequential Bobs (say, Bob$_{k}$). Alice and Bob$_{1}$ share an entangled state $\rho_{AB_{1}}$. Our aim is to demonstrate the sharing of nonlocality by multiple sequential observers. Since a projective measurement inevitably disturbs the system maximally; hence, in the sequential Bell test, if the first Bob (Bob$_{1}$) performs a sharp measurement, the entanglement is lost after the measurement. In such a case, no residual entanglement remains for the second sequential observer (Bob$_{2}$); consequently, the Bell inequality cannot be violated. If both the sequential observers obtain the sub-optimal quantum violations, then the first observer must have to perform an unsharp measurement.  

 In this work, the unsharp measurement corresponds to the noisy variant of projective measurements, i.e., the number of measurement operators is restricted to two. After performing the unsharp measurement, Bob$_{1}$ relays the post-measurement state to Bob$_{2}$, who performs an unsharp measurement intending to violate the Bell inequality. The chain runs up to arbitrary $k^{th}$ Bob (Bob$_{k}$) until the quantum violation of Bell's inequality is obtained. The $k^{\text{th}}$ Bob may perform the sharp measurement. It is quite known that in the standard scenario at most, two Bobs can sequentially demonstrate nonlocality through the violation of CHSH inequality \cite{silva2015}. We stress again that, before our work all the studies that demonstrated the sharing of nonlocality, the dimension of the system was assumed. 

We consider that Alice, upon receiving input $x=\{1,2\}$, always performs the projective measurements of observables $A_{1}$ and $A_{2}$. The CHSH form of Bell's inequality can be written as
	\begin{equation}
		\label{chsh}
		\mathcal{B}=\left(A_{1}+A_{2}\right) B_{1} +\left(A_{1}-A_{2}\right) B_{2} \leq 2.
	\end{equation}
	There are arbitrary $k$ number of sequential Bobs (say, Bob$_{k}$), who upon receiving input $y_{k}\in \{1,2\}$ perform measurements of  $B_{1}$ and $B_{2}$ producing outputs $b_{k}\in \{0,1\}$. We demonstrate how sequential quantum violations of CHSH inequality by multiple independent Bobs enable the DI certification of the unsharpness parameter. 
	
\begin{figure*}[ht]
				\label{Fig}
				\includegraphics[width=1.0\linewidth]{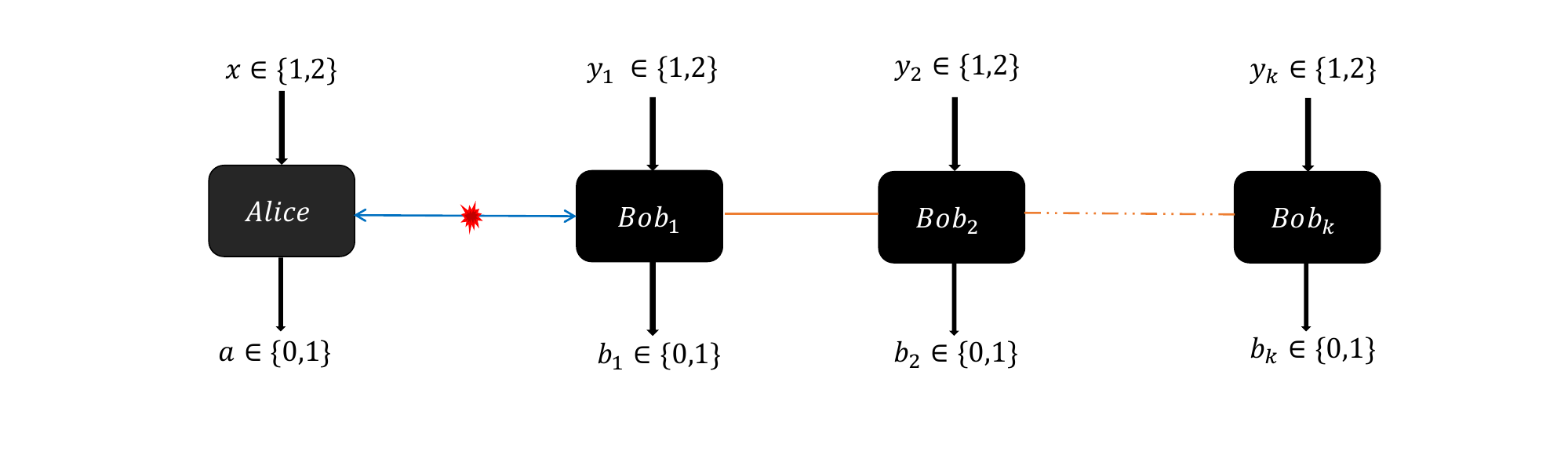}
				\caption{Black box diagram for sequential Bell test consisting of one Alice and multiple sequential Bobs (Bob$_{k}$). Alice shares an entangled state with Bob$_{1}$. After Bob$_{1} $'s unsharp measurement, the average state is relayed to Bob$_{2}$ and so on.}
			\end{figure*} 
			
Now, if Bob$_{k}$'s instrument is represented by measurement operators $\{K_{b_{k}|y_{k}}\}$ then after $(k-1)^{th}$ Bob's measurement, the average state shared between Alice and Bob$_{k}$ is 
	\ba
	\label{dp21}
	\rho_{AB_{k}}  = \frac{1}{2}\sum_{b_{k}\in \pm}\sum_{y_{k}=1}^{2} K_{b_{k}|y_{k}}^{\dagger}\ \rho_{AB_{(k-1)}}\ K_{b_{k}|y_{k}} \label{eq:D}
	\ea

	where $\forall k$,  $\sum_{b_{k}=\pm 1} K_{b_{k}|y_{k}}^{\dagger}K_{b_{k}|y_{k}}=\mathbb{I}$. Also, $K_{b_{k}|y_{k}}= \sqrt{E_{b_{k}|y_{k}}}$ where $E_{b_{k}|y_{k}}$ is the POVM. We consider that $\forall k$ and $\forall y_{k}$ the POVM \cite{busch,kunj14} is of the form 
 \ba
 E_{\pm|y_{k}}=\frac{1\pm\lambda_{k}}{2}\Pi_{y_k}^{+}+\frac{1\mp\lambda_{k}}{2}\Pi_{y_k}^{-}
 \ea
 where $\{\Pi_{y_k}^{\pm}\}$ are the projectors corresponding to the observable $B_{y_{k}}$ satisfying $\Pi_{y_k}^{+}+\Pi_{y_k}^{-}=\mathbb{I}$, and $\lambda_{k}$ is the unsharpness parameter for $k^{\text{th}}$ Bob. The measurement operators can then be written as, $K{\pm|y_{k}}= \alpha_{k} \mathbb{I} \pm \beta_{k} B_{y_{k}}$
	where 
	\begin{eqnarray}\alpha_{k} = \frac{1}{2}\Big(\sqrt{\frac{(1+\lambda_{k})}{2}}+ \sqrt{\frac{(1-\lambda_{k})}{2}}\Big)\\
\nonumber
\beta_{k} = \frac{1}{2}\Big(\sqrt{\frac{(1+\lambda_{k})}{2}}-\sqrt{\frac{(1-\lambda_{k})}{2}}\Big),
	\end{eqnarray}
	satisfying $\alpha_{k}^{2} +\beta_{k}^{2} = 1/2$ with $\alpha_{k}\geq \beta_{k}$.

We derive the maximum quantum value of CHSH expression for Alice and Bob$_{1}$  as
	\begin{align}
	\label{b1l}
		(	\mathcal{B}^{1})_Q=\lambda_{1}(	\mathcal{B})_Q^{opt}
	\end{align}
which is independent of assuming the dimension. Here, $\lambda_{1}$ is the unsharpness parameter of Bob$_{1}$.

	After his unsharp measurement, Bob$_1$ relays the average state to Bob$_2$. From Eq. (\ref{dp21}), the reduced state for Alice and Bob$_{2}$ can be written as
	\begin{eqnarray}
		\label{dp22}
		\nonumber
			\rho_{AB_{2}}&=&\dfrac{1}{2} \sum_{b_1 \in \{ +,-\}}\sum_{y_{1}=1}^{2} \left(\mathbb{I} \otimes K_{b_{1}|y_{1}}\right)  \rho_{AB_{1}}  \left(\mathbb{I} \otimes K_{b_{1}|y_{1}}\right)\\
				&=& 2\alpha^2_{1} \rho_{AB_{1}} + \beta^2_{1} \sum_{y_{1}=1}^{2}(\mathbb{I}\otimes B_{y_{1}})\rho_{AB_{1}} (\mathbb{I}\otimes B_{y_{1}}) .
	\end{eqnarray}	
For the time being, let us consider that each sequential Bob measures same set of observables $B_{1}$ and $B_{2}$, i.e., $\forall k, \ B_{y_{k=1}}\equiv B_{1}$ and $B_{y_{k=2}}\equiv B_{2}$. By using Eq. (\ref{dp22}), the maximum quantum value of CHSH expression for Alice and Bob$_{2}$ can be written as
	 
	 \ba
  \label{b2bell}
  \nonumber
	 	(\mathcal{B}^{2})_Q&=& max\Big(Tr[\rho_{AB_{2}} \mathcal{B}]\Big)\\
   &=&max\Bigg(Tr\Big[\rho_{AB_{1}}\Big((A_{1}+A_{2})\tilde{B}_{1}+(A_{1}-A_{2})\tilde{B}_{2}\Big)\Big]\Bigg)
	 \ea
	 where we assume Bob$_{2}$ performs sharp measurement. We derive the explicit forms of $\tilde{B}_{1}$ and $\tilde{B}_{2}$ as
  \ba
  \nonumber
\tilde{B}_{1}&=&(2\alpha_{1}^2+\beta_{1}^2)B_{1}+\beta_{1}^{2}B_{2}B_{1}B_{2}\\
\tilde{B}_{2}&=&(2\alpha_{1}^2+\beta_{1}^2)B_{2}+\beta_{1}^{2}B_{1}B_{2}B_{1}.
  \ea
Note that, Eq. (\ref{b2bell}) has a similar form of CHSH expression as in Eq. (\ref{chsh}), if $\tilde{B}_{1}$ and $\tilde{B}_{2}$ are considered to be effective observables of Bob$_{2}$. We can use the earlier SOS approach to obtain the maximum quantum value. However, $(\tilde{B}_{1})^{2}\neq \mathbb{I}$ and $(\tilde{B}_{2})^{2}\neq \mathbb{I}$, and hence they need to be properly normalized. By considering $\tilde{\omega}_{1}=||\tilde{B}_{1}||$ and $\tilde{\omega}_{2}=||\tilde{B}_{2}||$, and by using the SOS approach we have 
 \ba
 \label{maxchsh}
 (\mathcal{B}^{2})_{Q}= max\left(\omega_{1}\tilde{\omega}_{1} +\omega_{2}\tilde{\omega}_{2}\right).
 \ea
As we already proved earlier, to obtain the optimal quantum value Alice's and Bob's observables have to be mutually anticommuting. Hence, for Bob's (unnormalized) observables $\tilde{B}_{1}$ and $\tilde{B}_{2}$ we require,
\ba
\label{anti}
\{\tilde{B}_{1},\tilde{B}_{2}\}=4\alpha_{1}^2(\alpha_{1}^2+2\beta_{1}^{2}) \{B_{1},B_{2}\}+\beta_{1}^{4}\{B_{1},B_{2}\}^{3}=0 .
\ea
Since $\alpha_{1}>0$ and $\beta_{1}\geq 0$, Eq. (\ref{anti}) gives $\{B_{1},B_{2}\}=0$. In other words, the observables of Bob$_{2}$ have to be anticommuting to obtain the maximum quantum value of the Bell expression $(\mathcal{B}^{2})_{Q}$. This, in turn, provides 
\ba
\label{otilde}
\tilde{\omega}_{1}=\sqrt{(2\alpha_{1}^2)^{2}+(2\alpha_{1}^2+\beta_{1}^2)\beta_{1}^{2}\{B_{1},B_{2}\}^{2}}=2\alpha_{1}^2,
\ea
and $\tilde{\omega}_{2}=\tilde{\omega}_{1}=2\alpha_{1}^2$. Note that the above result is obtained by considering that Bob$_{1}$ and Bob$_{2}$ perform the measurements on the same set of observables $B_{1}$ and $B_{2}$. In Appendix \ref{appA}, we prove that to obtain the maximum quantum value $(\mathcal{B}^{2})_{Q}$ the choices of observables of Bob$_{2}$ have to be the same as Bob$_{1}$.

Using Eqs. (\ref{optbnn}) and  (\ref{otilde}), from Eq. (\ref{maxchsh}) we can then write
\begin{eqnarray}
\nonumber
    (\mathcal{B}^{2})_{Q} = 2\alpha_{1}^2 \ max\Big(\omega_{1}+\omega_{2}\Big)=2\alpha_{1}^2(\mathcal{B})_{Q}^{opt}.
\end{eqnarray}

Putting the value of $\alpha_{1}$, we write	$(\mathcal{B}^{2})_Q$ in terms of unsharpness parameter $\lambda_{1}$ of Bob$_{1}$ as
	\begin{equation}
		\label{llbell}
		(\mathcal{B}^{2})_Q=\dfrac{1}{2}\Big(1+\sqrt{1-\lambda_{1}^2}\Big)(\mathcal{B})_Q^{opt} .
	\end{equation}
	
Hence, we have derived the maximum quantum values of Bell expressions for two sequential Bobs where $(\mathcal{B})_Q^{opt}$ is common, but they differ by the coefficients, which are solely dependent only on $\lambda_1$. Using Eq. (\ref{b1l}), the quantum value of the CHSH expression $(\mathcal{B}^{2})_Q$ in Eq. (\ref{llbell}) can be written in terms of $(\mathcal{B}^{1})_Q$ as 
\begin{equation}
\label{toff}
  	(\mathcal{B}^{2})_Q=\sqrt{2}\Big(1+\sqrt{1-\Bigg(\frac{	(\mathcal{B}^{1})_Q}{(\mathcal{B})_{Q}^{opt}}\Bigg)^{2}}\Big)  .
\end{equation}

This implies that if  $	(\mathcal{B}^{1})_Q$  increases, then $(\mathcal{B}^{2})_Q$ decreases, i.e., the more the Bob$_{1}$ disturbs the system, the more he gains the information, and consequently the quantum value $(\mathcal{B}^{2})_Q$ of Bob$_{2}$ decreases. Hence, there is a trade-off between $(\mathcal{B}^{2})_Q$ and $(\mathcal{B}^{1})_Q$, which eventually form an optimal pair demonstrating the certification of unsharpness parameter.

\begin{figure}[h]
	\label{fig2}
		\centering 
		{ \includegraphics[width=8cm,height=8cm]{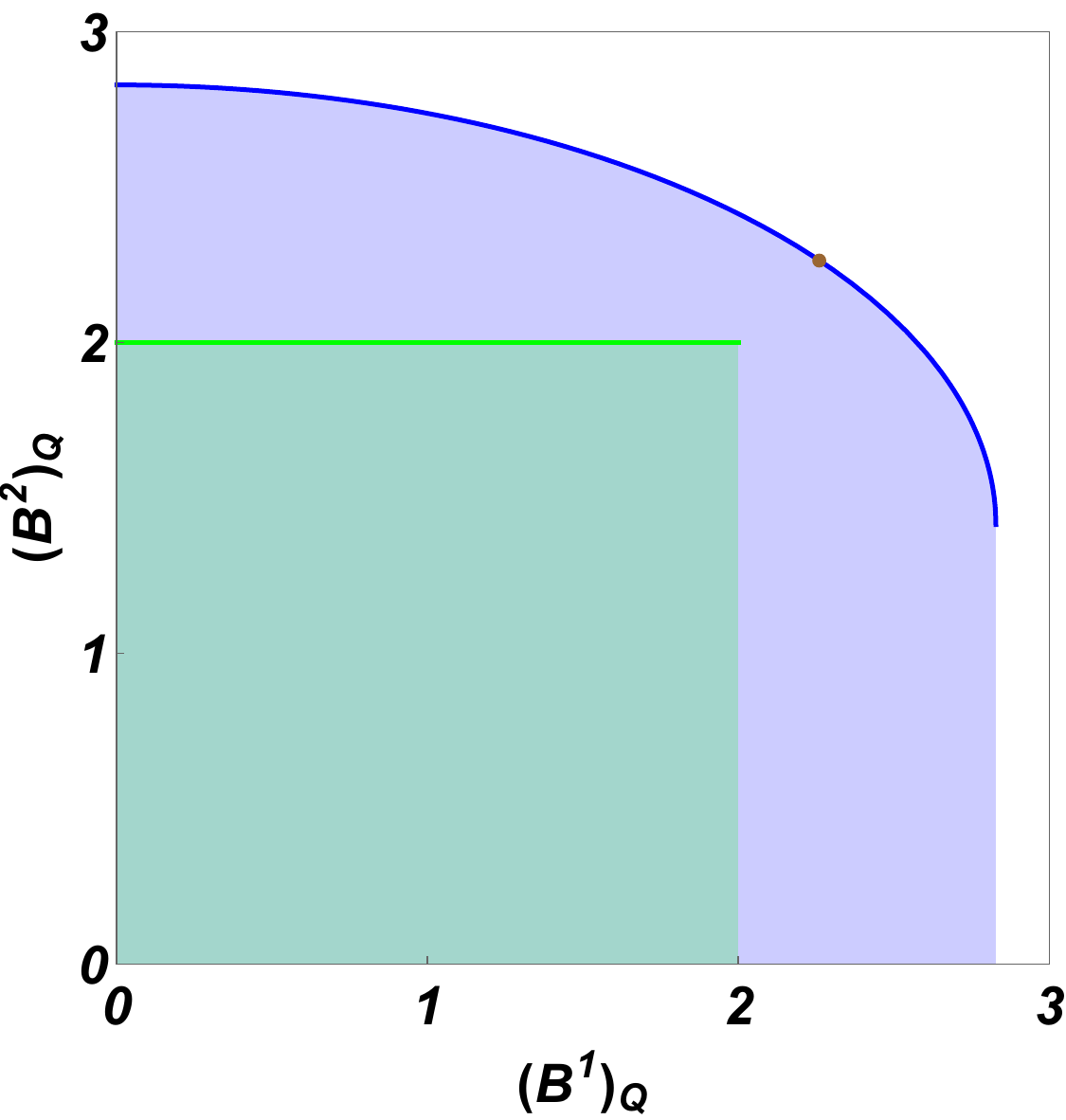}}
		\caption{\footnotesize \centering Optimal trade-off between quantum bound of CHSH inequality of Bob$_1$ and Bob$_2$ is shown by the solid blue curve while the
shaded portion gives the suboptimal range. The solid green line is for
classical bound of CHSH inequality for the same two observers.}
	\end{figure}
	
In \figurename{ 2}, we plot the optimal trade-off characteristics between $(\mathcal{B}^{1})_Q$ and $(\mathcal{B}^{2})_Q$. The green line corresponds to the maximum classical value of the CHSH expression where  Bob$_{2}$ can get a maximum value independent of Bob$_{1}$, i.e., there is no trade-off in classical theory. The blue curve exhibits the trade-off between quantum values of $(\mathcal{B}^{2})_Q$ and $(\mathcal{B}^{1})_Q$ where each point on it certifies a unique value of unsharpness parameter $\lambda_{1}$. For example, when $(\mathcal{B}^{2})_Q=(\mathcal{B}^{1})_Q$, the value of $\lambda_1=4/5=0.80$ is certified, as shown in the figure by brown dot.

Note that in our protocol we consider Alice always performs sharp measurements.  Before making the self-testing statements of our protocol, let us examine whether our protocol certifies sharp measurements of Alice. As mentioned earlier that the blue curve in \figurename{ 2} represents the optimal trade-off between $(\mathcal{B}^{2})_Q$ and $(\mathcal{B}^{1})_Q$. If Alice performs unsharp measurement then the trade-off curve will always be below the blue curve. For instance, $(\mathcal{B}^{2})_Q=	(\mathcal{B}^{1})_Q=8\sqrt{2}/5$ (represented by the brown dot over the blue line) can never be reached unless Alice performs the sharp measurement of her anti-commuting observables. The same argument holds for any point over the blue curve in \figurename{ 2}. We are now in the position to make the DI self-testing statements of our protocol.

\subsection{DI self-testing statements}
The sub-optimal quantum values $(\mathcal{B}^{1})_Q$ and $(\mathcal{B}^{2})_Q$ form an optimal pair $\{(\mathcal{B}^{1})_Q, (\mathcal{B}^{2})_Q\}$ that uniquely certifies the shared state between Alice and Bob$_{1}$, the set of observables, and the unsharpness parameter $\lambda_{1}$. The self-testing statements are the following,

(i) Alice performs sharp measurements of two mutually anticommuting observables on her local subsystem in any arbitrary local dimension.   

(ii) Bob$_{1}$ performs unsharp measurement corresponding to two observables which are also mutually anticommuting in any arbitrary local dimension. The set of observables for Bob$_{1}$ and Bob$_{2}$ are the same.

(iii) Alice and Bob$_{1}$ share a maximally entangled state in any arbitrary dimension.

(iv) The optimal pair $\{(\mathcal{B}^{1})_Q, (\mathcal{B}^{2})_Q\}$ self-tests the unsharpness parameter $\lambda_1$ which in turn certifies the shared entangled state between Alice, Bob$_{1}$ and Bob$_{2}$. In \figurename{ 2}, each point on the surface of blue curve certifies a unique value of unsharpness parameter $\lambda_{1}$.

\subsection{Robust certification of unsharpness parameter}
Note, however, that the experimental implementation of any protocol inevitably introduces noise and imperfections. We provide an argument to demonstrate how our certification protocol is robust to the noise. In the real experimental scenario, the maximum values of CHSH expressions $(\mathcal{B}^{2})_Q$ and $(\mathcal{B}^{1})_Q$ may not be achieved, and
hence unique certification of $\lambda_{1}$ may not be accurate. In such a case, we can certify the range within which $\lambda_{1}$ can belong. 

 The quantum advantage for Bob$_{1}$ requires $(\mathcal{B}^{1})_Q  > 2$, which fixes the $(\lambda_{1})_{min}=1/\sqrt{2}\approx0.707 $ as $(\mathcal{B})_{Q}^{opt}=2\sqrt{2}$. Thus, any value of $\lambda_{1} \in [1/\sqrt{2},1]$ provides quantum advantage for Bob$_{1}$. However, this range has to be further restricted if the nonlocality is extended to Bob$_{2}$. 
   To obtain advantage for Bob$_{2}$, two sequential Bob requires $(\mathcal{B}^{1})_Q, (\mathcal{B}^{2})_Q>2$ and in turn necessitates $\lambda_{2}=1$.
   From Eq. (\ref{llbell}), we get
   \begin{align}
   \label{l1max}
       \lambda_{1} < \sqrt{1-\Bigg(\frac{2(\mathcal{B}^{2})_{Q}}{(\mathcal{B})_{Q}^{opt}}-1\Bigg)^{2}}
   \end{align}
   which in turn fixes the upper bound of $(\lambda_{1})_{max} = \sqrt{2(\sqrt{2}-1)}\approx 0.912$. Hence, when both  Bob$_{1}$  and  Bob$_{2}$ get quantum advantage, the interval $0.707<\lambda_{1}< 0.912$ is certified. 
   
   Now, as an example, let us consider an experiment in which we attempt to certify the desired value of $\lambda_{1}=0.74$ corresponding to the optimal pair $\{(\mathcal{B}^{1})_Q,(\mathcal{B}^{2})_Q\}$.  In such a scenario, we need the value of $\{(\mathcal{B}^{1})_Q,(\mathcal{B}^{2})_Q\}\approx\{2.093,2.365\}$. However, due to imperfections, accurate quantum values may not be obtained. Instead, experimentalist gets  $\{(\mathcal{B}^{1})_Q,(\mathcal{B}^{2})_Q\}\approx\{2.05,2.34\}$. In such a case, we cannot obtain the desired value of $\lambda_{1}$ but will deduce a range of $\lambda_{1}$ within which it has to be confined. From Eqs. (\ref{b1l}) and (\ref{l1max}), we can calculate the range of $\lambda_{1}$ as $0.724<\lambda_{1}<0.755$. Thus, depending upon the observed quantum values, the range of $\lambda_{1}$ can be confined, i.e., the more perfect the experimental determination of $\{(\mathcal{B}^{1})_Q,(\mathcal{B}^{2})_Q\}$, the certified range of $\lambda_{1}$ becomes narrower.

 \subsection{Sharing of nonlocality by the third Bob}
 We examine whether the sharing of nonlocality can be extended to the third Bob (Bob$_{3}$). If so, we can certify two unsharpness parameters. Also, the range of $\lambda_{1}$ can be made more restrictive if the third Bob can share the nonlocality. By using Eq. (\ref{dp21}), the maximum quantum value of CHSH expression $(\mathcal{B}^{3})_Q= max(Tr[\rho_{AB_{3}} (\mathcal{B})])$ between Alice and Bob$_{3}$ can be calculated  where $\rho_{AB_{3}}$ is the average state shared between Alice and Bob$_{3}$. Then, 
 \begin{eqnarray}
 \nonumber
	(\mathcal{B}^{3})_Q&=&max\left(Tr\left[\Big(2\alpha^2_{2} \rho_{AB_{2}} + \beta^2_{2} \sum_{y_2=1}^{2}(\mathbb{I}\otimes B_{y_2})\rho_{AB_{2}} (\mathbb{I}\otimes B_{y_2})\Big)  \mathcal{B}\right]\right)\\
\end{eqnarray}
which can be re-written in a similar form as in Eq. (\ref{b2bell}), is given by
\begin{eqnarray}
\label{b3bell}
(\mathcal{B}^{3})_Q=max\Bigg(Tr\Big[\rho_{AB_{1}}\Big((A_{1}+A_{2})\tilde{\tilde{B}}_{1}+(A_{1}-A_{})\tilde{\tilde{B}}_{2}\Big)\Big]\Bigg).
\end{eqnarray} 
Here, $\tilde{\tilde{B}}_{1}$ and $\tilde{\tilde{B}}_{2}$ represent effective observables of Bob$_{3}$ are derived as
\begin{eqnarray}
\nonumber
\tilde{\tilde{B}}_{i}&=&(4\alpha^2_{1}\alpha_2^{2}+2\beta_{1}^{2}\beta_2^{2})B_{i}+(2\alpha_2^{2}\beta_{1}^{2}+2\alpha_{1}^{2}\beta_{2}^{2})(B_{i}+B_{j}B_{i}B_{j})\\
&+&\beta_{1}^{2}\beta_2^{2}(B_{j}B_{i}B_{j}+B_{i}B_{j}B_{i}B_{j}B_{i})
\end{eqnarray}
with $i(j)_{i\neq j} \in \{1,2\}$. Again $(\tilde{\tilde{B}}_{i})^{2}\neq \mathbb{I}$ and hence $\tilde{\tilde{B}}_{i}$'s need to be properly normalized. Following the earlier argument, we can use the aforesaid  SOS approach to obtain the optimal value of $(\mathcal{B}^{3})_{Q}$. By considering $\tilde{\tilde{\omega}}_{1}=||\tilde{\tilde{B}}_{1}||$ and $\tilde{\tilde{\omega}}_{2}=||\tilde{\tilde{B}}_{2}||$, and by using the SOS approach we obtain  
 \ba
 \label{bell3}
 (\mathcal{B}^{3})_{Q}= max\left(\omega_{1}\tilde{\tilde{\omega}}_{1} +\omega_{2}\tilde{\tilde{\omega}}_{2}\right) .
 \ea

 Note that, to obtain the maximum quantum value, $\tilde{\tilde{B}}_{1}$ and $\tilde{\tilde{B}}_{2}$ have to be mutually anticommuting, i.e., 
\begin{eqnarray}
\nonumber
&&\{\tilde{\tilde{B}}_{1},\tilde{\tilde{B}}_{2}\}=\Big(8(4 \alpha^2_{1} \alpha^4_{2} \beta^2_{1} +2\alpha^4_{1} (\alpha^4_{2} + 2 \alpha^2_{2} \beta^2_{1}) -3 \beta^4_{1} \beta^4_{1})\Big) \{B_{1},B_{2}\}\\
\nonumber
&+&\Big(4 (\alpha_{2}^{4} \beta_{1}^{4} + 
   2 \alpha_{2}^{2} \beta_{1}^{2} (2\alpha_{1}^{2} + \beta_{1}^{2}) \beta_{2}^{2} + \alpha_{1}^{2} (\alpha_{1}^{2} + 2\beta_{1}^{2}) \beta_{2}^{4})\Big)\{B_{1},B_{2}\}^{3}\\
\nonumber
&+&(\beta_{1}^{2}\beta_2^{2})^{2}\{B_{1},B_{2}\}^{5}=0
\end{eqnarray}
which provides $\{B_{1},B_{2}\}=0$. Then, Bob$_{3}$ also requires the anticommuting observables to obtain the maximum quantum value. We then calculate
\begin{eqnarray}
\nonumber
\tilde{\tilde{\omega}}_{1}&=&\Bigg(16\alpha^4_{1}\alpha_2^{4}+\Big(4\alpha_2^{4}\beta_{1}^{4}+2\alpha_{1}^{2}\alpha_2^{2}\beta_{1}^{2}(\alpha_2^{2}+2\beta_{2}^{2})+\alpha^4_{1}(2\alpha_{2}^{2}\beta_{2}^{2}+\beta_{2}^{4})\Big)\\
&&\{B_{1},B_{2}\}^{2}+\Big(2\alpha_2^{2}\beta_{1}^{4}\beta_{2}^{2}+\beta_{1}^{2}(2\alpha_1^{2}+\beta_{1}^2)\beta_{2}^4\Big)\{B_{1},B_{2}\}^{4}\Bigg)^{1/2} .
\end{eqnarray}
Since, $\{B_1,B_2\}=0$, we obtain $\tilde{\tilde{\omega}}_{1}=4\alpha^2_{1}\alpha_2^{2}$, and similarly we find  $\tilde{\tilde{\omega}}_{2}=\tilde{\tilde{\omega}}_{1}=4\alpha^2_{1}\alpha_2^{2}$.

Then, the maximum quantum value $(\mathcal{B}^{3})_Q$ in Eq. (\ref{bell3}) becomes
\begin{eqnarray}
  \nonumber
    (\mathcal{B}^{3})_{Q}= 4\alpha^2_{1}\alpha_2^{2} \ max\Big(\omega_{1}+\omega_{2}\Big)=4\alpha^2_{1}\alpha_2^{2}(\mathcal{B})_{Q}^{opt} .
\end{eqnarray}

Putting the values of $\alpha_{1}$ and $\alpha_{2}$, we write $(\mathcal{B}^{3})_Q$ in terms of unsharpness parameters of Bob$_{1}$ and Bob$_{2}$ as
	\begin{equation}
		\label{bell3q}
		(\mathcal{B}^{3})_Q=\dfrac{1}{4}\Big(1+\sqrt{1-\lambda_{1}^2}\Big)\Big(1+\sqrt{1-\lambda_{2}^2}\Big)(\mathcal{B})_Q^{opt} .
	\end{equation}

 Now, to exhibit the quantum violation for three sequential Bobs, we have to show that $(\mathcal{B}^{1})_Q,(\mathcal{B}^{2})_Q,(\mathcal{B}^{3})_Q>2$. For $(\mathcal{B}^{1})_Q>2$, the lower bound on $\lambda_{1}$ is $\lambda_{1} > 1/\sqrt{2}$. From Eq. (\ref{llbell}), we can calculate the lower bound of $\lambda_{2}$ for $(\mathcal{B}^{2})_Q>2$ is
	 \begin{align}
       \lambda_2 > \frac{4}{(\mathcal{B})_{Q}^{opt}\Big(1+\sqrt{1-\Big(\frac{2}{(\mathcal{B})_{Q}^{opt}}}\Big)^2}= \dfrac{2}{\sqrt{2}+1} \approx 0.828 .
   \end{align}

	By considering that Bob$_1$ and Bob$_2$ implement their unsharp measurements with minimum required values of unsharpness parameters at their respective sites, and by
	substituting the values of $(\lambda_{1})_{min}\approx0.707$ and $(\lambda_{2})_{min}\approx0.828 $ in Eq. (\ref{bell3q}), we get $(\mathcal{B}^{3})_Q=1.89$. This means Alice and Bob$_{3}$ cannot violate CHSH inequality when both Bob$_{1}$ and Bob$_{2}$ violate it. 

\section{Elegant Bell inequality and its local and preparation non-contextual bounds}
We extend the sequential sharing of nonlocality by using another well-known Bell inequality known as Gisin's elegant Bell inequality \cite{gisin}. The elegant Bell expression can be written as,
		\begin{eqnarray}
			\label{ebi}
			\nonumber
			\mathbb{E}&=&(A_{1}+A_{2}+A_{3}-A_{4})\otimes B_{1}+(A_{1}+A_{2}-A_{3}\\
			&+&A_{4}) \otimes B_{2}+(A_{1}-A_{2}+A_{3}+A_{4})\otimes B_{3}.
		\end{eqnarray} 
whose local bound is $(\mathbb{E})_{l} \leq 6$. We show that the elegant Bell expression has two bounds, the local bound (which can be considered as trivial preparation non-contextual bound) and a nontrivial preparation non-contextual bound when there exists a relational constraint between the observables of Alice and Bob. We show that the non-trivial preparation non-contextual bound $(\mathbb{E})_{pnc} \leq 4$. 

Before proceeding further, we briefly introduce the notion of preparation non-contextuality in an ontological model of quantum theory. Consider that a preparation procedure $P$ prepares a density matrix $\rho$, and a measurement procedure $M$ realizes the measurement of a POVM $E_{k}$. Quantum theory predicts the probability of obtaining a particular outcome $k$ is $p(k|P, M)=Tr[\rho E_{k}]$ - the Born rule. In an ontological model of quantum theory, preparation procedures assign a probability distribution  $\mu_{P}(\lambda|\rho)$ on ontic states $\lambda \in \Lambda$ where $\Lambda$ is the ontic state space. Given the measurement procedure, the ontic state $\lambda$ assigns a response function $\xi_{M}(k|\lambda,E_{k})$.  A viable ontological model must reproduce the Born rule, i.e., $\forall k, \rho, E_{k}:\ \ p(k|P, M)=\int_{\Lambda} \mu_{P}(\lambda|\rho) \xi_{M}(k|\lambda,E_{k})d\lambda$. 

The dependencies of $P$ and $M$ do not appear if the ontological model is preparation and measurement non-contextual, respectively. Two preparation procedures $P$ and $P^{\prime}$ are said to be operationally equivalent if they can not be distinguished by any measurement, implying, $\forall k,M:\hspace{.1cm}p(k|P, M)=p(k|P^{\prime},M) $. In quantum theory, such preparation procedures are realized by the density matrix $\rho$. Such equivalence in operational theory can be reflected in the ontic state level, assuming preparation non-contextuality \cite{spek05}. 
An ontological model of an operational quantum theory is considered to be preparation non-contextual if two preparation procedures $P$ and $P^{\prime}$ prepare the same density matrix $\rho$, and no measurement can operationally distinguish the context by which $\rho$ is prepared, i.e., $\forall k, M :\hspace{.1cm} p(k|P,M) = p(k|P^{\prime},M) \Rightarrow  \forall \lambda : \mu(\lambda|\rho,P) = \mu(\lambda|\rho,P^{\prime})$, implying two ontic state distributions are equivalent irrespective of the contexts $ P $ and $ P^{\prime}$ \cite{spek05,spekk09,hameedi17b,ghorai18,pan19}.
	
Let us intuitively understand the notion of preparation non-contextuality in the  CHSH scenario. Consider that Alice and Bob share an entangled state $\rho_{AB}$ and Alice measures $A_{1}$ and $A_{2}$. Alice's measurements on her local system produces density matrices $\rho_{A_{1}}$ and $\rho_{A_{2}}$ on Bob's side corresponding to  measurement contexts $A_{1}$ and $A_{2}$, respectively. The non-signaling condition demands that $\rho_{A_{1}}$ and $\rho_{A_{2}}$ cannot be distinguishable by any measurement of Bob, i.e., $\rho_{A_{1}}=\rho_{A_{2}}\equiv \sigma$. Equivalently, in an ontological model we assume that $\mu(\lambda|\sigma, A_{1})=\mu(\lambda|\sigma, A_{2})$, i.e., the distribution of ontic states are preparation non-contextual. Intuitively, preparation non-contextuality implies the locality assumption in the Bell scenario. In other words, every probability distribution that violates a Bell inequality can also be regarded as proof of preparation contextuality as proved in \cite{uola}. Here, we provide a modified version of the proof.

 For this, by using Bayes' theorem we write the joint probability distribution in the ontological model as 
\begin{equation}
\label{jp}
 p(a, b|A_{i}, B_{j}) = \sum_{\lambda}p(a|A_{i}, B_{j})p(\lambda|a,A_{i})p(b|B_{j}, \lambda).
\end{equation}
Now, the no-signaling condition implies the marginal probability of Alice's side is independent of Bob's input and hence we can write
\begin{equation}
\label{jpd}
 p(a, b|A_{i}, B_{j}) = \sum_{\lambda}p(a|A_{i})p(\lambda|a,A_{i})p(b|B_{j}, \lambda).
\end{equation}
Using Bayes' theorem we can write $p(a|A_{i})p(\lambda|a,A_{i})=\mu(\lambda|A_{i})p(a|\lambda,A_{i})$ where we specifically denoted the probability distribution $p(\lambda|A_{i})$ as $\mu(\lambda|A_{i})$.

From Eq. (\ref{jpd}), we then obtain
\begin{equation}
\label{}
 p(a, b|A_{i}, B_{j}) = \sum_{\lambda}\mu(\lambda|A_{i})p(a|\lambda,A_{i})p(b|B_{j}, \lambda).
\end{equation}

Due to the assignment of the same ontic-state distribution for the different preparation procedures on Bob's side, the assumption of preparation non-contextuality is enforced. Then the preparation non-contextual assumption for Bob's preparation reads as $\mu(\lambda|\rho,{A_{1}})=\mu(\lambda|\rho,{A_{2}})\equiv\mu(\lambda)$, which in turn provides
\begin{equation}
\label{}
 p(a, b|A_{i}, B_{j}) = \sum_{\lambda}\mu(\lambda)p(a|\lambda,A_{i})p(b| \lambda, B_{j}).
\end{equation}
which is the desired factorizability condition commonly derived for a local hidden variable model. 
Therefore, we can argue that whenever in an ontological model $p(a, b|A_{i},B_{j})$ satisfies preparation non-contextuality, this, in turn, satisfies locality in a hidden variable model.

 Here we go one step further. The above case involving the CHSH scenario is a trivial one. We introduce a nontrivial form of preparation non-contextuality in the Bell experiment involving more than two inputs. Consider a Bell experiment where Alice receives four inputs and performs the measurements of four observables $A_{1}$, $A_{2}$, $A_{3}$ and $A_{4}$. Bob receives three inputs and performs the measurement of three observables $B_1$ , $B_2$ and $B_3$.  In quantum theory, when Alice and Bob share an entangled state $\rho_{AB}$
 \ba
 \label{p34}
 \rho_{a|A_{i}}+\rho_{a\oplus 1|A_{i}}=\rho_{a|A_{i^{\prime}}}+\rho_{a\oplus 1|A_{i^{\prime}}}\equiv \sigma
 \ea
 where $i,i^{\prime}=1,2,3,4$ with $i\neq i^{\prime}$, and $ \rho_{a|A_{i}}=Tr_{A}\left[\rho_{AB}\Pi_{A_{i}}\otimes \mathbb{I}\right]$. This is within the premise of preparation non-contextuality in an ontological model as the distribution of ontic states is assumed to be equivalent for those two preparation procedures $i$ and $i^{\prime}$. As argued above, such a trivial preparation non-contextuality can be attributed to the locality in a Bell experiment. In that case, the local bound of the elegant Bell expression $(\mathbb{E})_{l}\leq6$.
 
 We introduce a nontrivial form of preparation non-contextuality in an ontological model of quantum theory by imposing an additional relational constraint on Alice's measurement observables. For this, we consider that the joint probability satisfies  
 \ba
\forall b, j, \   &&P(a, b|A_{1},B_{j})+\sum\limits_{i=2}^{4} P(a\oplus 1, b|A_{i},B_{j})\\
 \nonumber
 &=&P(a\oplus 1, b|A_{1},B_{j})+\sum\limits_{i=2}^{4} P(a, b|A_{i},B_{j})
 \ea
 which in quantum theory implies that
 \ba 
 \label{pp}
  \nonumber
 &&\rho_{a\oplus 1|A_{1}}+\rho_{a |A_{2}}+\rho_{a|A_{3}}+\rho_{a|A_{4}} \\
 &=& \rho_{a|A_{1}}+\rho_{a\oplus 1 |A_{2}}+\rho_{a\oplus 1|A_{3}}+\rho_{a\oplus 1|A_{4}}
 \ea 
 Since $ \rho_{a|A_{i}}=Tr_{A}\left[\rho_{AB}\Pi_{A_{i}} \otimes \mathbb{I}\right]$ with $\Pi_{A_{i}}=\left(\mathbb{I}+A_{i}\right)/2$, the above Eq. ({\ref{pp}}) implies that $A_{1}-A_{2}-A_{3}-A_{4}=0$. Along with the equivalence in Eq. (\ref{p34}) a non-trivial  constraint Eq. (\ref{pp}) is also imposed on Alice's preparation procedures. In such a case, the local bound
reduces to the preparation non-contextual bound $(\mathbb{E})_{l}\leq 4$ as $A_{1}=A_{2}+A_{3}+A_{4}$. Thus, the quantum violation of it provides a weaker notion of nonlocality, which we call nontrivial preparation contextuality. It is well-known that steering is a weaker form of nonlocality, but the relation between steering and nontrivial preparation contextuality needs to be properly explored. While we plan to do it in another occasion, interested reader may see a relevant work \cite{uola}. However, for our present purpose it is not directly relevant and hence we skip such discussion.  
 
Note that the quantum value of $(\mathbb{E})_{Q}$ has also to be calculated by considering this constraint. Below we show that the optimal quantum value $(\mathbb{E})_{Q}^{opt}=4\sqrt{3}$ satisfies the constraint $A_{1}=A_{2}+A_{3}+A_{4}$.

\section{Sharing preparation contextuality and certification of multiple unsharpness parameters}	
 
We derive the optimal quantum value $(\mathbb{E})_{Q}^{opt}$ without using the dimension of the system. For this, we again use the SOS approach developed in Sec. II. We define a positive semidefinite operator $\langle\chi\rangle_Q \geq 0$ so that $\langle\chi\rangle_Q +\mathbb{E}= \Omega_3$ where $\Omega_3$ is a positive quantity. By considering suitable positive operators $L_1$, $L_2$ and $L_3$ we can write
\begin{align}
	\label {g31}
	\chi =\frac{1}{2} \left( \omega_{1} L_1^\dagger L_1+\omega_{2} L_2^\dagger L_2+\omega_{3} L_3^\dagger L_3\right) 
\end{align}
where $\omega_{1}$, $\omega_{2}$ and $\omega_{3}$ are positive numbers that will be determined soon. 
For our purpose, we choose
\begin{align}
	\label{li}
	\nonumber
	L_{1}|\psi\rangle_{AB}=\frac{A_{1}+A_{2}+A_{3}-A_{4}}{\omega_{1}} |\psi\rangle_{AB} -B_{1} |\psi\rangle_{AB}\\
	L_{2}|\psi\rangle_{AB}=\frac{A_{1}+A_{2}-A_{3}+A_{4}}{\omega_{2}} |\psi\rangle_{AB} -B_{2} |\psi\rangle_{AB}\\
	\nonumber
	L_{3}|\psi\rangle_{AB}=\frac{A_{1}-A_{2}+A_{3}+A_{4}}{\omega_{3}} |\psi\rangle_{AB} -B_{3} |\psi\rangle_{AB}
\end{align}
where $\omega_{1}=||(A_{1}+A_{2}+A_{3}-A_{4})|\psi\rangle_{AB}||_{2}, 	\omega_{2}=||(A_{1}+A_{2}-A_{3}+A_{4})|\psi\rangle_{AB}||_{2}$ and $	\omega_{3}=||(A_{1}-A_{2}+A_{3}+A_{4})|\psi\rangle_{AB}||_{2}$. Substituting Eq. (\ref{li}) in Eq. (\ref{g31}), we get
\begin{equation}
    \langle\chi\rangle_Q=-(\mathbb{E})_Q+\sum_{i=1}^{3}\omega_{i} .
\end{equation}
Hence, the optimal value $(\mathbb{E})_Q^{opt}$ is obtained if
$\langle\chi\rangle_Q = 0$, i.e.,
\begin{equation}
    (\mathbb{E})_Q^{opt}=max(\omega_{1}+\omega_{2}+\omega_{3})
\end{equation}
where
\begin{eqnarray}
\label{O1}
\nonumber
 \omega_1= \sqrt{4+\langle\{A_{1},(A_{2}+A_{3}-A_{4})\}+\{A_{2},(A_{3}-A_{4})\}-\{A_{3},A_{4}\}\rangle}\\
\end{eqnarray}
and similarly for  $\omega_2$ and $\omega_3$.

We use the concavity inequality $\sum_{i=1}^{n}\omega_{i}\leq \sqrt{n\sum_{i=1}^{n}(\omega_{i})^{2}}$ where the equality holds when $\omega_{i}$'s are equal to each other.  We can then write
\begin{equation}
\label{inq}
    (\mathbb{E})_{Q}\leq max\Big(\sqrt{3(\omega_{1}^{2}+\omega_{2}^{2}+\omega_{3}^{2})}\Big).
\end{equation}
Now, by using the expressions of $\omega_{1}$, $\omega_{2}$ and $\omega_{3}$ we can write
\begin{eqnarray}
\omega_{1}^{2}+\omega_{2}^{2}+\omega_{3}^{2}={_{AB}\bra{\psi}}(12+\delta)\ket{\psi}_{AB}
\end{eqnarray}
where 
\ba
\nonumber
\delta=(\{A_{1},(A_{2}+A_{3}+A_{4})\}-\{A_{2},(A_{3}+A_{4})\}-\{A_{3},A_{4}\}).\\
\ea
To maximize $\delta$, we consider that there exist a state $\ket{\psi^{\prime}}$ which can be written as $\ket{\psi^{\prime}}=(A_{1}-A_{2}-A_{3}-A_{4})\ket{\psi}_{AB}$ such that $\ket{\psi}_{AB}\neq 0$.
We can show that $\langle{\psi^{\prime}}|{\psi^{\prime}}\rangle=4-\langle\delta\rangle$. By rearranging we can write $\langle\delta\rangle=4-\langle{\psi^{\prime}}|{\psi^{\prime}}\rangle$. Clearly, the maximum value $\langle\delta\rangle_{max}=4$ is obtained when $\langle{\psi^{\prime}}|{\psi^{\prime}}\rangle=0$. Since, $\ket{\psi}_{AB}\neq 0$, we find the condition of maximization
\ba
\label{cons}
A_{1}-A_{2}-A_{3}-A_{4}=0.
\ea
Since, $\langle\delta\rangle_{max}=4$, we find $max(\omega_{1}^{2}+\omega_{2}^{2}+\omega_{3}^{2})=16$. From Eq. (\ref{inq}) we can then write $(\mathbb{E})_{Q}\leq4\sqrt{3}$.

Using Eq. (\ref{cons}), a few steps of calculations gives the following relations that need to be satisfied by Alice's observables given by
\begin{eqnarray}
\label{Aac1}
\{A_{1},A_{2}\}=\{A_{1},A_{3}\}=\{A_{1},A_{4}\}=\frac{2}{3},\\
\label{Aac2}
\{A_{2},A_{3}\}=\{A_{2},A_{4}\}=\{A_{3},A_{4}\}=-\frac{2}{3}.
\end{eqnarray}
Putting Eqs. (\ref{Aac1}) and (\ref{Aac2}) in Eq. (\ref{O1}), we get $\omega_{1}= 4/\sqrt{3}$. A similar calculation gives us $\omega_{2}=\omega_{3}= 4/\sqrt{3}$. 
Since, $\omega_{1}=\omega_{2}=\omega_{3}$, the optimal quantum value of elegant Bell expression is $(\mathbb{E})_{Q}^{opt}= 4\sqrt{3}$. Note here again that we have not imposed any bound on the dimension of the system throughout the derivation.

As mentioned, the optimal quantum value of $\mathbb{E}$ for any given
dimension is obtained when $\langle\chi\rangle_Q = 0$, implying that 
\ba
\forall i\in [3],\  L_{i}\ket{\psi}_{AB}=0.
\ea
Further, using the conditions in Eq. (\ref{li}), Bobs observables  $B_{1}$, $B_{2}$, and $B_{3}$ can be written as
\begin{eqnarray}
\nonumber
B_{1}&=&\frac{\sqrt{3}(A_{1}+A_{2}+A_{3}-A_{4})}{4},\\ 
\nonumber
B_{2}&=&\frac{\sqrt{3}(A_{1}+A_{2}-A_{3}+A_{4})}{4}, \\ 
B_{3}&=&\frac{\sqrt{3}(A_{1}-A_{2}+A_{3}+A_{4})}{4}.
\end{eqnarray}
It is straightforward to show that Bob's observables satisfy,
\begin{eqnarray}
\nonumber
\{B_{1},B_{2}\}&=&\frac{\sqrt{3}}{4}\{(A_{1}+A_{2}+A_{3}-A_{4}),(A_{1}+A_{2}-A_{3}+A_{4})\}\\
&=&\frac{\sqrt{3}}{4}\Big(\{A_{1},A_{2}\}+\{A_{3},A_{4}\}\Big).
\end{eqnarray}
Putting the values of Alice's anticommuting relations from Eqs. (\ref{Aac1}) and (\ref{Aac2}), we get $\{B_{1},B_{2}\}=0$.
Similarly, it can also be shown that $\{B_{1},B_{3}\}=0$ or $\{B_{2},B_{3}\}=0$. Thus, Bobs observables  $B_{1},B_{2}$ and $B_{3}$ have to be mutually anticommuting to obtain the optimal quantum value $(\mathbb{E})_{Q}^{opt}=4\sqrt{3}$. Following the argument developed for CHSH, it can be proved that the shared state between Alice and Bob has to be a maximally entangled state as in Eq. (\ref{mentngled}).
 \subsection{Certification of multiple unsharpness parameters}
		Similar to the sequential CHSH scenario, for elegant Bell inequality, the average state shared between Alice and Bob$_{k}$ if Alice and Bob$_{(k-1)}$ shares a maximally entangled state and each Bob performs a dichotomic POVM measurement in the sequential scheme is
		\ba
		\label{g22}
		\rho_{AB_{k}}  = \frac{1}{3}\sum_{b_{k}\in \pm}\sum_{y_{k}=1}^{3} K_{b_{k}|y_{k}}\rho_{AB_{(k-1)}}K_{b_{k}|y_{k}} \label{eq:D}
		\ea
		where $\{K_{b_{k}|y_{k}}\}$ are the Kraus operators satisfying $\sum_{b_{k}} K_{b_{k}|y}^{\dagger}K_{b_{k}|y_{k}}=\mathbb{I}$ and $K_{b_{k}|y_{k}}= \sqrt{E_{b_{k}|y_{k}} }$.

		The quantum value of elegant Bell expression due to the unsharp measurement of Bob$_{1}$ irrespective of the dimension is calculated as
	\begin{align}
	\label{e1opt}
		(	\mathbb{E}^{1})_Q=\lambda_{1}(	\mathbb{E})_Q^{opt}.
	\end{align}
 
		Let us assume that every sequential Bob measures same set of observables $B_{1}$, $B_{2}$ and $B_{3}$ i.e., $\forall k, \ B_{y_k=1}\equiv B_{1}$ and so on. The average post-measurement state after Bob$_{1} $'s unsharp measurement can be written as
		\begin{eqnarray}
			\nonumber
			\rho_{AB_{2}}&=&\dfrac{1}{3} \sum_{b_{1}\in \{ +,-\}}^{}\sum_{y_{1}=1}^{3} \left(\mathbb{I} \otimes K_{b_{1}|y_{1}}\right)  \rho_{AB_{1}}  \left(\mathbb{I} \otimes K_{b_{1}|y_{1}}\right)\\
\label{dp32}
			&=&\dfrac{1}{3} \Big[ 6\alpha^2_{1} \rho_{AB_{1}} +2 \beta^2_{1} \sum_{y_{1}=1}^{3}(\mathbb{I}\otimes B_{y_{1}})\rho_{AB_{1}}(\mathbb{I}\otimes B_{y_{1}})\Big] .
		\end{eqnarray} 
		
		Using $\rho_{AB_{2}}$ in Eq. (\ref{dp32}), the maximum quantum value of elegant Bell expression between Alice and Bob$_{2}$ is given by
	
		\begin{eqnarray}
		\nonumber
			(\mathbb{E}^{2})_Q &=& max\Big(Tr[\rho_{AB_{2}} \mathbb{E}]\Big)\\
			\nonumber
			\label{ebi2}
				&=&max\Big(Tr\Big[\rho_{AB_{1}}\Big((A_{1}+A_{2}+A_{3}-A_{4})B_{1}^{\prime}+(A_{1}+A_{2}\\
				&-&A_{3}+A_{4})B_{2}^{\prime}+(A_{1}-A_{2}+A_{3}+A_{4})B_{3}^{\prime}\Big)\Big]\Big).
		\end{eqnarray}
 We derive $B_{1}^{\prime}$, $B_{2}^{\prime}$ and $B_{3}^{\prime}$ as
  \begin{eqnarray}
  \label{bprime}
  \nonumber
B_{1}^{\prime}&=&2(\alpha_{1}^2+\frac{\beta_{1}^2}{3})B_{1}+\frac{2}{3}\beta_{1}^2(B_{2}B_{1}B_{2}+B_{3}B_{1}B_{3}),\\
 	B_{2}^{\prime}&=&2(\alpha_{1}^2+\frac{\beta_{1}^2}{3})B_{2}+\frac{2}{3}\beta_{1}^2(B_{1}B_{2}B_{1}+B_{3}B_{2}B_{3}),\\
 	\nonumber
 	B_{3}^{\prime}&=&2(\alpha_{1}^2+\frac{\beta_{1}^2}{3})B_{3}+\frac{2}{3}\beta_{1}^2(B_{1}B_{3}B_{1}+B_{2}B_{3}B_{2}).
 \end{eqnarray}
 
Eq. (\ref{ebi2}) has a complete resemblance with elegant Bell inequality in Eq. (\ref{ebi}) with the effective observables of Bob are $B_{1}^{\prime}$, $B_{2}^{\prime}$ and $B_{3}^{\prime}$. However, $(B_{i}^{\prime})^{2}\neq \mathbb{I}$ with $i\in [3]$ and hence they need to be normalized. By assuming, $\omega_{i}^{\prime}=||B_{i}^{\prime}||$ and using the SOS approach we get 
 \ba
 \label{max3}
			(\mathbb{E}^{2})_{Q}= max \left(\omega_{1}\omega_{1}^{\prime}+\omega_{2}\omega_{2}^{\prime}+\omega_{3}\omega_{3}^{\prime}\right).
 \ea
Optimization of quantum value of $(\mathbb{E}^{2})_{Q}$ demands Bob's observables $B_{1}^{\prime}$, $B_{2}^{\prime}$ and $B_{3}^{\prime}$ have to be mutually anticommuting. It is explicitly shown in Appendix \ref{appB} that the anticommutation $\{B_{i}^{\prime},B_{j}^{\prime}\}$ can be written in terms of $\{B_{i},B_{j}\}$ where $i(j) \in 1,2,3$ with $i\neq j$. It is proved that $\{B_{i}^{\prime},B_{j}^{\prime}\}=0$ implying $\{B_{i},B_{j}\}=0$ as the parameters $\alpha_1$ and $\beta_1$ are positive.

 In other words, Bob$_{2}$ also requires the anticommuting observables to obtain the maximum quantum value of $(\mathbb{E}^{2})_{Q}$. We then calculate
\begin{eqnarray}
\nonumber
\omega_{1}^{\prime}&=&\Bigg((2\alpha_{1}^2+\frac{2}{3}\beta_{1}^2)^{2}+(\frac{2}{3}\beta_{1}^2)^{2}\Big(4+B_{2}\{B_{1},B_{2}\}B_{3}\{B_{1},B_{3}\}\\
\nonumber
&-&B_{2}\{B_{1},B_{2}\}B_{1}-B_{1}B_{3}\{B_{1},B_{3}\}+B_{3}\{B_{1},B_{3}\}B_{2}\{B_{1},B_{2}\}\\
\nonumber
&-&B_{3}\{B_{1}B_{3}\}B_{1}-B_{1}B_{2}\{B_{1},B_{2}\}\Big)+(2\alpha_{1}^2+\frac{2}{3}\beta_{1}^2)\frac{2}{3}\beta_{1}^2\\
&&\Big(\{B_{1},B_{2}\}^{2}+\{B_{1},B_{3}\}^{2}-4\Big)\Bigg)^{1/2}.
\end{eqnarray}
Using  $\{B_{i},B_{j}\}=0$, we get $\omega_{1}^{\prime}=2(\alpha_{1}^2-\frac{\beta_{1}^2}{3})$. Similarly, we find  $\omega_{2}^{\prime}=\omega_{3}^{\prime}=\omega_{1}^{\prime}$.

Putting the values of $\omega_{1}^{\prime}$, $\omega_{2}^{\prime}$ and $\omega_{3}^{\prime}$, the maximum quantum value of the elegant Bell expression $(\mathbb{E}^{2})_Q$ can be obtained from Eq. (\ref{max3}) as
\begin{eqnarray}
\nonumber
(\mathbb{E}^{2})_Q&=&2(\alpha_{1}^2-\frac{\beta_{1}^2}{3})(\mathbb{E})_Q^{opt}.
\end{eqnarray}
 	
	By inserting the values of $\alpha_{1}$ and $\beta_{1}$, we have	$(\mathbb{E}^{2})_Q$ in terms of unsharpness parameter as
	\begin{equation}
		\label{ele2q}
		(\mathbb{E}^{2})_Q=\dfrac{1}{3}\Big(1+2\sqrt{1-\lambda_{1}^2}\Big)(\mathbb{E})_Q^{opt}.
	\end{equation}
	
			\begin{figure}[h]
	\label{fig2}
		\centering 
		{ 
		{\includegraphics[width=8cm,height=8cm]{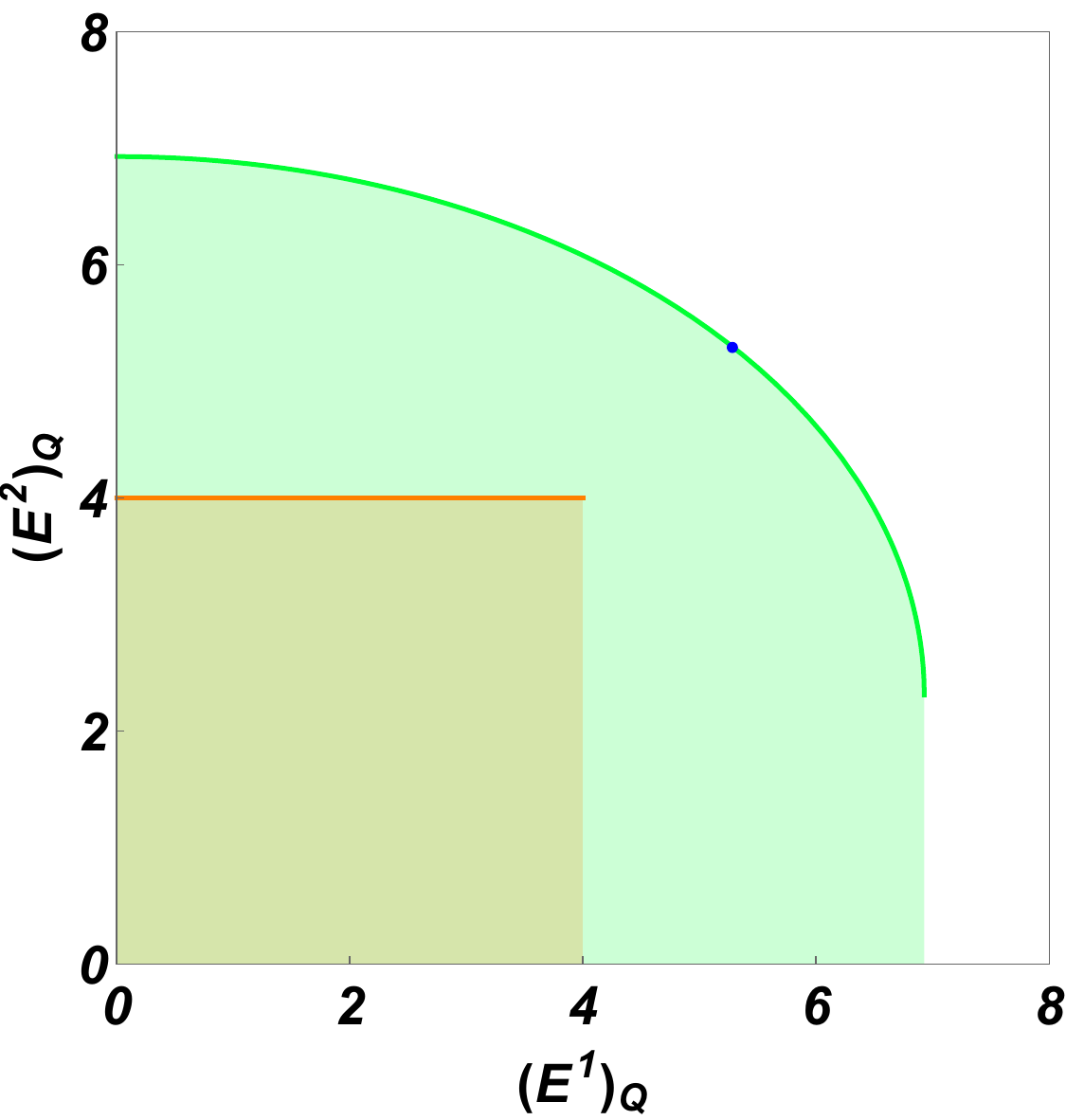}}}
		\caption{\footnotesize \centering Optimal trade-off between quantum bound of elegant Bell inequality of Bob$_1$ and Bob$_2$ is shown by the solid green curve while the
shaded portion gives the suboptimal range. The solid orange line is for
classical bound for the same two observers.}
	\end{figure}
		
		Note that, the maximum values of $(\mathbb{E}^{1})_Q$ and $(\mathbb{E}^{2})_Q$ in Eq. (\ref{e1opt}) and Eq. (\ref{ele2q}) are dependent only on the unsharpness parameter $\lambda_{1}$ as $(\mathbb{E})^{opt}_Q$ is common. Writing $(\mathbb{E}^{2})_Q$ in terms of $(\mathbb{E}^{1})_Q$ we get
\begin{equation}
\label{e2opt}
  	(\mathbb{E}^{2})_Q=\frac{4}{\sqrt{3}}\Big(1+2\sqrt{1-\Bigg(\frac{	(\mathbb{E}^{1})_Q}{(\mathbb{E})^{opt}_Q}}\Bigg)^{2}\Big)  .
\end{equation}

It can be seen for Eq. (\ref{e2opt}) that if Bob$_{1}$ extracts more information then the quantum value of $	(\mathbb{E}^{1})_Q$  increases and consequently decreases the value of $(\mathbb{E}^{2})_Q$ and hence there is a trade-off between the values of $(\mathbb{E}^{2})_Q$ and $(\mathbb{E}^{1})_Q$. \figurename{ 3} represents the optimal trade-off relation written in Eq. (\ref{e2opt}) where the green curve shows the quantum values where each point on its surface certifies a unique value of unsharpness parameter $\lambda_{1}$. For $(\mathbb{E}^{1})_Q=(\mathbb{E}^{2})_Q=12(4+\sqrt{3})/13\approx5.291$, both the Bob$_{1}$ and Bob$_{2}$ gets equal advantage as shown by the blue point on the surface of the curve which uniquely certifies the value of sharpness parameter $\lambda_1=(\sqrt{57+24\sqrt{3}})/13\approx0.763$.
   
	Similarly, the quantum value of elegant Bell expression between Alice and Bob$_{3}$ can be calculated as $(\mathbb{E}^{3})_Q= Tr[\rho_{AB_{3}} (\mathbb{E})]$ where $\rho_{AB_{3}}$ is the average state shared between Alice and Bob$_{3}$. Then, the elegant Bell expression between Alice and Bob$_{3}$ can be re-written as
\begin{eqnarray}
\label{ebi3}
\nonumber
(\mathbb{E}^{3})_Q&=&Tr\Big[\rho_{AB_{1}}\Big((A_{1}+A_{2}+A_{3}-A_{4})B_{1}^{\prime\prime}+(A_{1}+A_{2}\\
&-&A_{3}+A_{4})B_{2}^{\prime\prime}+(A_{1}-A_{2}+A_{3}+A_{4})B_{3}^{\prime\prime}\Big)\Big]
\end{eqnarray}
where $B^{\prime\prime}_{1}$, $B^{\prime\prime}_{2}$ and $B^{\prime\prime}_{3}$ are explicitly defined in Appendix \ref{appC}.

Now, we can use the SOS approach as mentioned earlier to obtain the optimal value of $(\mathbb{E}^{3})_{Q}$. But, $(B^{\prime\prime}_{i})^{2}\neq \mathbb{I}$ and hence needs to be normalised.  By considering $\omega^{\prime\prime}_{1}=||B^{\prime\prime}_{1}||$, $\omega^{\prime\prime}_{2}=||B^{\prime\prime}_{2}||$  and $\omega^{\prime\prime}_{3}=||B^{\prime\prime}_{3}||$ and by using the SOS approach we obtain  
 \ba
 (\mathbb{E}^{3})_{Q}= max\left(\omega_{1}\omega^{\prime\prime}_{1} +\omega_{2}\omega^{\prime\prime}_{2}++\omega_{3}\omega^{\prime\prime}_{3}\right).
 \ea

 Note that, $B^{\prime\prime}_{i}$ and $B^{\prime\prime}_{j}$ with $i(j)_{i\neq j} \in \{1,2,3\}$ can again be proved to be mutually anticommuting. That is, Bob$_{3}$ requires the anticommuting observables to obtain the optimal quantum value. Using the anticommutation relation we then calculate $\omega^{\prime\prime}_{1}=4(\alpha_{1}^2-\frac{\beta_{1}^2}{3})(\alpha_{2}^2-\frac{\beta_{2}^2}{3})$. Similarly, we find  $\omega^{\prime\prime}_{2}=\omega^{\prime\prime}_{3}=\omega^{\prime\prime}_{1}$.

The quantum value for Alice and Bob$_{3}$ can be written as
\begin{eqnarray}
    (\mathbb{E}^{3})_{Q} &=& 4(\alpha_{1}^2-\frac{\beta_{1}^2}{3})(\alpha_{2}^2-\frac{\beta_{2}^2}{3}) \ max\Big(\omega_{1}+\omega_{2}+\omega_{3}\Big)\\
    &=&4(\alpha_{1}^2-\frac{\beta_{1}^2}{3})(\alpha_{2}^2-\frac{\beta_{2}^2}{3})(\mathbb{E})_{Q}^{opt}.
\end{eqnarray}

Writing $(\mathbb{E}^{3})_{Q}$ in terms of the unsharpness parameters $\lambda_{1}$ and $\lambda_{2}$ we get
	\begin{equation}
		\label{ele3q}
		(\mathbb{E}^{3})_Q=\dfrac{1}{9}\Big(1+2\sqrt{1-\lambda_{1}^2}\Big)\Big(1+2\sqrt{1-\lambda_{2}^2}\Big)(\mathbb{E})_Q^{opt}
	\end{equation}
which can be generalised for Alice and any arbitrary $k^{\text{th}}$ Bob (Bob$_{k}$) as
		\begin{equation}
		\label{kbobe}
		(\mathbb{E}^{k})_Q=\dfrac{1}{3^{k}}\prod_{i=1}^{k-1}\sqrt{3}\Big(1+2\sqrt{1-\lambda_{i}^2}\Big)(\mathbb{E})_Q^{opt}.
	\end{equation}
	
	Further, $(\mathbb{E}^{3})_Q$ can be simplified in term of $(\mathbb{E}^{1})_Q$, $(\mathbb{E}^{2})_Q$ as
		\begin{equation}
		\label{tradeoff33}
		(\mathbb{E}^{3})_Q=\dfrac{1}{9}(1+\Delta_{1})\Bigg(1+\sqrt{\frac{4+4\Delta_{1}-\Delta_{2}}{1+\Delta_{1}}}\Bigg)
	\end{equation}
	where $\Delta_{1}=2\sqrt{1-\Big(\frac{(\mathbb{E}^{1})_{Q}}{(\mathbb{E})^{opt}_Q}\Big)^{2}}$ and $\Delta_{2}=\dfrac{\sqrt{3}(\mathbb{E}^{1})_Q}{4}$.

	\begin{figure}[ht]
	\label{fig3}
		\centering 
		{{\includegraphics[width=1.0\linewidth]{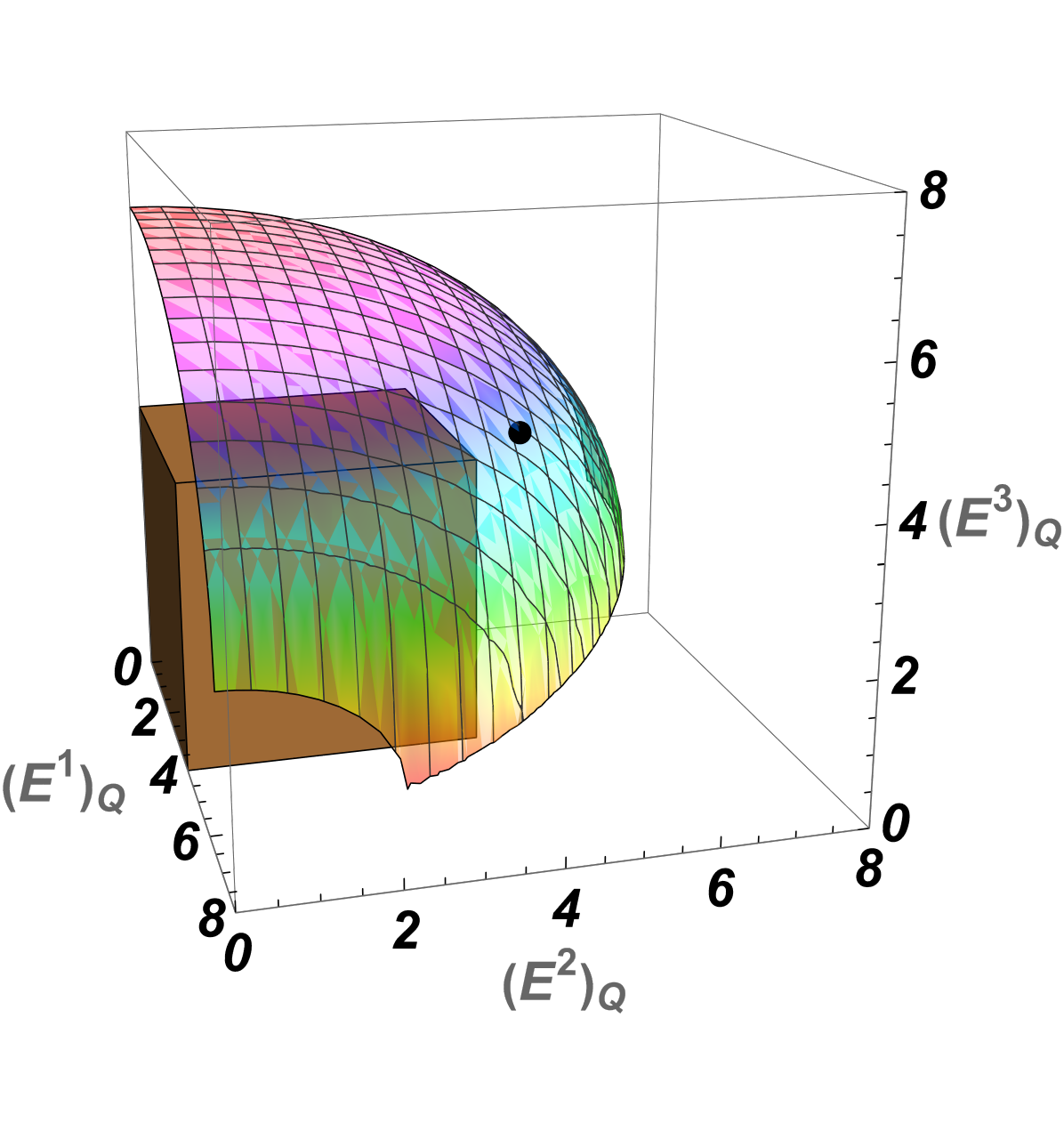}}}
		\caption{\footnotesize \centering Optimal trade-off between quantum bound of elegant Bell inequality of Bob$_1$, Bob$_{2}$ and Bob$_3$. The black point on the three-dimensional graph indicates the point which certifies the unsharpness parameters $\lambda_{1}$ and $\lambda_{2}$ when quantum values of all three sequential Bobs are considered to be equal.}
	\end{figure}
	
	Optimal trade-off between quantum bound of elegant Bell inequality of Bob$_1$, Bob$_{2}$ and Bob$_3$ are given by Eq. (\ref{tradeoff33}) and plotted in \figurename{ 4}. The brown cube represents the preparation non-contextual bound showing no trade-off. The three-dimensional semi-paraboloid over the cube represents the trade-off between quantum bound. For sharp measurement of Bob$_{3}$ with $\lambda_{3}=1$, each point on the surface of the semi paraboloid in \figurename{ 4} uniquely certifies $\lambda_{1}$ and $\lambda_{2}$. The black point on the surface of the semi paraboloid uniquely certifies $\lambda_{1}=0.644$ and $\lambda_{2}=0.763$ for $(\mathbb{E}^{1})_Q=(\mathbb{E}^{2})_Q=(\mathbb{E}^{3})_Q =4.462$.
\subsection{DI self-testing statement in sequential scenario}

Thus, the optimal triple $\{(\mathbb{E}^{1})_Q, (\mathbb{E}^{2})_Q$,  $(\mathbb{E}^{3})_Q \}$ uniquely certifies the states shared between Alice and of Bob$_{1}$,  Bob$_{2}$ and Bob$_3$, their observables, and the unsharpness parameters of Bob$_{1}$ and Bob$_{2}$. The self-testing statements are the following,

(i) Alice performs the sharp measurement of her observables in any arbitrary local dimension, satisfying the functional relations in Eq. (\ref{cons}).   

(ii) Bob$_{1}$ performs unsharp measurement corresponding to
three observables that are mutually anticommuting in any arbitrary local dimension. The set of observables of  Bob$_{2}$ and  Bob$_{3}$ are the same.

(iii) Alice and Bob$_{1}$ share a maximally entangled state in any arbitrary dimension.

(iv) The optimal triple pair $\{(\mathbb{E}^{1})_Q,(\mathbb{E}^{2})_Q,(\mathbb{E}^{3})_Q\}$ certifies the unsharpness parameters $\lambda_{1}$ and $\lambda_{2}$ which, in turn, self-tests the shared state between Alice, Bob$_{1}$, Bob$_{2}$ and Bob$_{3}$ respectively. As shown in \figurename{ 4}, each point on the surface of semi-paraboloid self-tests  unique values of unsharpness parameters $\lambda_{1}$ and $\lambda_{2}$.

	\subsection{Robust certification of unsharpness parameters}
	
Similar to the sequential CHSH scenario, for sub-optimal violation of elegant Bell expression, the certifiable range of $\lambda_1$ from Eqs. (\ref{e1opt}) and (\ref{ele2q}) can be calculated as
	
   \begin{align}
   \label{rangel1}
   \frac{(\mathbb{E}^{1})_Q}{(\mathbb{E})^{opt }_Q}<\lambda_{1}<\sqrt{1-\frac{1}{4}\Bigg(\frac{3(\mathbb{E}^{2})_Q}{\lambda_{2}(\mathbb{E})^{opt }_Q}-1\Bigg)^{2}}.
   \end{align}
   Eq. (\ref{rangel1}) shows that for the optimal quantum value of $(\mathbb{E})_Q^{opt}$, the lower bound of $\lambda_{1}$ depends only on the maximum quantum value of $(\mathbb{E}^{1})_Q$.
   Since, $(\mathbb{E})_Q^{opt}$ is $4\sqrt{3}$, for violating the elegant Bell inequality Bob$_{1}$ requires $(\mathbb{E}^{1})_Q>4$, which fixes lower bound of unsharpness parameter as $(\lambda_{1})_{min} \approx 1/\sqrt{3}\approx 0.57$. Any value of $\lambda_{1} \in [1/\sqrt{3},1]$  therefore provides the quantum advantage for Bob$_{1}$. 
   
   The upper bound of unsharpness parameter $\lambda_{1}$ is a function of $\lambda_{2}$ and the maximum quantum value of $(\mathbb{E}^{2})_Q$. To get advantage for Bob$_{2}$, two sequential Bob requires $(\mathbb{E}^{1})_Q, (\mathbb{E}^{2})_Q>4$. Considering Bob$_{2}$ performs sharp measurement ($\lambda_{2}=1$), the upper bound on the unsharpness parameter i.e., $(\lambda_{1})_{max}$ from Eq. (\ref{rangel1}) is calculated as  $(\lambda_{1})_{max} =\sqrt{\sqrt{3}/2}\approx 0.93$. Thus, when both Bob$_{1}$ and Bob$_{2}$ get quantum advantage, the interval $0.57<\lambda_{1}< 0.93$ can be certified. 
   
   Now, the range of $\lambda_{1}$ becomes more restricted if the quantum advantage is further extended to Bob$_{3}$ which demands $(\mathbb{E}^{1})_Q, (\mathbb{E}^{2})_Q, \text{and}\  (\mathbb{E}^{3})_Q>4$. In such a case, from Eqs. (\ref{e1opt}) and (\ref{ele3q}), the interval of $\lambda_{1}$ is narrower down to $0.57 < \lambda_{1} <  0.77 $ which requires a more efficient experimental realization. 
   
	Again, if sequential Bobs up to Bob$_{3}$ get the quantum advantage, the range of $\lambda_2$ can be calculated from Eqs. (\ref{ele2q}) and (\ref{ele3q}) as
	\begin{eqnarray}
	\label{lam2}
	\dfrac{3(\mathbb{E}^{2})_Q}{\xi_{1}(\mathbb{E})_Q^{opt}}<\lambda_2<\sqrt{1-\dfrac{1}{4}\Bigg(\dfrac{9(\mathbb{E}^{3})_Q}{\xi_{1}}-1\Bigg)^{2}}
	\end{eqnarray}
	where $\xi_{1}=\Big(1+2\sqrt{1-\lambda_{1}^{2}}\Big)$. Eq. (\ref{lam2}) shows that both the upper and lower
bound of $\lambda_{2}$ are dependent on $\lambda_{1}$. Thus, the range  $0.65 < \lambda_{2} <  0.87 $ can be certified. Further, from Eq. (\ref{ele3q}) the lower bound of $(\lambda_{3})_{min} $ is calculated as,
\begin{equation}
    (\lambda_3)_{min}=\dfrac{9(\mathbb{E}^{3})_Q}{\xi_{1} \Big(1+2\sqrt{1-\lambda_{2}^{2}}\Big)(\mathbb{E})_Q^{opt}}\approx 0.78 .  
\end{equation}
	
	Now, considering Bob$_1$, Bob$_{2}$ and Bob$_3$ implement their unsharp measurements with lower critical values of unsharpness parameters at their respective sites, from Eq. (\ref{kbobe}) we get $(\mathbb{E}^{4})_Q=3.84$, i.e., Alice and Bob$_{4}$ cannot violate elegant Bell inequality, thereby will not provide any quantum advantage over preparation non-contextual bound.

  \section{Summary and discussion}

The precise control of quantum devices plays a crucial role in the development of quantum technologies. Therefore, developing elegant protocols for certifying quantum devices is indispensable in quantum information theory. In this work, we provided DI self-testing of unsharp instruments based on two Bell inequalities, viz., the CHSH inequality and the elegant Bell inequality. Note that the optimal quantum violation of a Bell inequality uniquely self-tests the state and the observables. We discussed that such a violation does not certify the post-measurement states, and hence self-testing the unsharp instrument is not possible in that way. We demonstrated that the sequential Bell test by multiple independent observers on one side of the Bell experiment enables the self-testing of post-measurement states, which, in turn, uniquely self-tests the unsharp instruments. The optimal quantum violation of CHSH inequality can be achieved only for sharp measurement, and for unsharp measurement, one obtains the sub-optimal value. Notably, the sub-optimal quantum value may also arise due to the nonideal preparation of the state or the inappropriate choices of the local observables. However, as explicitly discussed in \cite{tava20exp}, the primary criticism regarding DI self-testing of unsharp measurement through a Bell test arises from Naimark’s theorem, which states that any non-projective measurement can be viewed as a projective measurement in a higher-dimensional Hilbert space. Since there is no bound on dimension in DI self-testing, one can always argue that the sub-optimal quantum violation arises from the inappropriate choice of observables in higher-dimension and not from the unsharp measurement. 

By providing the dimension-independent optimization of the sequential Bell test, we provided a scheme for self-testing of unsharp instruments. As mentioned, the sequential sharing of nonlocality by multiple independent observers plays a crucial role in our work. Crucially, such a sharing of quantum correlation without assuming the dimension of the system has not hitherto been discussed. Here, we impose no bound on the dimension of the system, and the quantum devices are taken as black boxes. We introduced an elegant SOS approach enabling us to derive the maximum quantum values in the sequential Bell test. We note here that the semi-DI certification of the unsharp instruments in a sequential prepare-measure scenario was demonstrated in \cite{mohan} by using a qubit system. In \cite{mohan}, the authors leave DI self-testing of an unsharp instrument as an open question which is now provided in our work.  

We first considered the sequential sharing of CHSH nonlocality where Alice always performs sharp measurement and an arbitrary $k$ number of sequential Bobs (Bob$_{k}$) who perform unsharp measurements on their local sub-system. If the first observer (Bob$_{1}$) performs a sharp projective measurement, the entanglement between Alice and Bob$_{1}$ will be lost, and there is no chance that Alice and Bob$_{2}$ will violate the CHSH inequality. On the other hand, if Bob$_{1}$ performs an unsharp measurement, a sufficient residual entanglement may remain to exhibit the violation of CHSH inequality between Alice and Bob$_{2}$. In such a case, we jointly maximized the sub-optimal quantum advantages for both sequential observers and demonstrated that there exists a trade-off relation between the two sequential quantum violations. The sub-optimal quantum violations form an optimal pair which eventually self-tests the unsharp instrument of Bob$_{1}$ along with the entangled state and the observables of Alice, Bob$_{1}$ and Bob$_{2}$. In the CHSH scenario, at most, two sequential Bobs can share the nonlocality, and hence only one unsharpness parameter can be self-tested. 

We extend our treatment to the sharing of preparation contextuality based on the elegant Bell inequality. We demonstrated that the preparation contextuality can be shared up to a maximum of three sequential Bobs. We jointly maximized the sub-optimal sequential quantum values by using the SOS approach without assuming the dimension of the system. There is a trade-off between the three sub-optimal quantum values for Bob$_{1}$, Bob$_{2}$, and Bob$_{3}$  violating the elegant Bell inequality. Three sub-optimal quantum values form an optimal triple which in turn self-tests the unsharp instruments of Bob$_{1}$ and Bob$_{2}$. In the process of maximization of the sub-optimal values, the entangled state and the observables of Alice and of Bob$_{1}$, Bob$_{2}$, and Bob$_{3}$ are self-tested. 

We note that due to unavoidable losses and imperfection in the actual experimental scenarios, the unique certification of the unsharpness parameter is not possible. In such a case, we provided an analysis of robust certification so that a range of the unsharpness parameter can be certified. The more perfect the actual experiment, one can achieve more accurate the certification of the unsharpness parameters. 

We conclude by proposing the following future direction and potential applications. Our work can be further generalized to self-test an arbitrary number of unsharp instruments based on the quantum violation of a family of preparation non-contextual inequalities proposed in \cite{ghorai18}. It can be interesting to study the sequential sharing of preparation contextuality by multiple numbers of observers on one side of the Bell experiment. Using the SOS approach one can simultaneously maximize the sub-optimal quantum values corresponding to the independent sequential observers. This can be an interesting line of future study.  Our work has immediate application in generating a higher amount of DI-certified randomness. In a recent work, based on the sequential sharing of nonlocality by using the two-qubit entangled state, the generation of a higher amount of randomness was proposed \cite{ran1}, which is experimentally tested in \cite{val21}. The self-testing protocol provided here can be used to generate a higher amount of DI-certified randomness. This calls for further study.
  \section*{Acknowledgment}PR acknowledges the support from the research grant DST/ICPS/QuEST/2019/4. AKP acknowledges the support from the research grant MTR/2021/000908.
  
  		\begin{widetext}
\appendix
\section{Calculation to show that every sequential Bob requires same set of observables}
\label{appA}
	Here we show that the set of observables require for achieving the maximum quantum violation for Bob$_{1}$ and Bob$_{2}$ are the same. Let Bob$_{1}$ performs measurement of $B_{1}$ and $B_{2}$ on his local subsystems upon receiving input $y_{1}\in \{1,2\}$ on the entangled state $\rho_{AB_{1}}$, and Bob$_{2}$ upon receiving input $y_{2}\in \{1,2\}$ performs measurements of the observables $B_{3}$ and $B_{4}$ on a state $\rho_{AB_{2}}$ producing outputs $b_{2}\in \{0,1\}$.
	Putting $\rho_{AB_{2}}$ from Eq. (\ref{dp22}), we can explicitly write the CHSH expression between Alice and Bob$_{2}$ as
	\begin{eqnarray}
	\label{a1}
		(\mathcal{B}^{2})_Q&=&Tr\Big[\rho_{AB_{1}}\left(\Big(A_{1}+A_{2}\Big) \bar{B}+\Big((A_{1}-A_{2})\bar{\bar{B}}\right)\Big]
	\end{eqnarray}
	where	$\bar{B}=2\alpha^2_{1}B_{3}+\beta^2_{1}\Big(B_{1}B_{3}B_{1}+B_{2}B_{3}B_{2}\Big)$ and $\bar{\bar{B}}=2\alpha^2_{1}B_{4}+\beta^2_{1}\Big(B_{1}B_{4}B_{1}+B_{2}B_{4}B_{2}\Big)$.

	By writing Eq. (\ref{a1}) in terms of the unsharpness parameter and rearranging, the CHSH expression can be written as 
		\begin{eqnarray}
		\label{ap1}
		(\mathcal{B}^{2})_Q&=&Tr\Bigg[\rho_{AB_{1}}\Bigg(\gamma_{1}\Big((A_{1}+A_{2}) B_{3}+(A_{1}-A_{2})B_{4}\Big)+\dfrac{\gamma_{1}^{\prime}}{2}\Big((A_{1}+A_{2}) (B_{1}B_{3}B_{1}+B_{2}B_{3}B_{2})+(A_{1}-A_{2})(B_{1}B_{4}B_{1}+B_{2}B_{4}B_{2}\Big)\Bigg)\Bigg]
	\end{eqnarray}
	where $\gamma_{1}=\dfrac{1}{2}({1+\sqrt{1-\lambda_{1}^{2}}})$ and $\gamma_{1}^{\prime}=\dfrac{1}{2}({1-\sqrt{1-\lambda_{1}^{2}}})$.
	
The first term in Eq. (\ref{ap1}) has the form of CHSH inequality where $B_{3}$ and $B_{4}$ can be proved as mutually anticommuting by the SOS approach.
Since $\gamma_{1} > \gamma_{1}^{\prime}$, we can then write
\begin{equation}
    	(\mathcal{B}^{2})_Q\leq\gamma_{1}\  max\Big(||(A_{1}+A_{2})|| +||(A_{1}-A_{2})||\Big)+\dfrac{\gamma_{1}^{\prime}}{2}\Big((A_{1}+A_{2}) (B_{1}B_{3}B_{1}+B_{2}B_{3}B_{2})+(A_{1}-A_{2})(B_{1}B_{4}B_{1}+B_{2}B_{4}B_{2})\Big).
\end{equation}
It is simple to show that $\{(B_{1}B_{3}B_{1}+B_{2}B_{3}B_{2}),(B_{1}B_{4}B_{1}+B_{2}B_{4}B_{2})\}=0$ which requires $B_{1}=B_{3}$ and $B_{2}=B_{4}$,
and in turn $\{B_{1},B_{2}\}=0$.
	
	\section{Detailed calculation for sequential quantum value of elegant Bell expression for Alice and Bob$_{2}$}
	\label{appB}
 	To derive optimal quantum value of the elegant Bell expression written in Eq. (\ref{ebi2}) we require the (unnormalized) observables to be $B_{1}^{\prime}$, $B_{2}^{\prime}$ and $B_{3}^{\prime}$ mutually anticommuting. We show that $\{B_{i}^{\prime}, B_{j}^{\prime}\}=0$ implies $\{B_{i}, B_{j}\}=0$ where $i,j=1,2,3$ with $i\neq j$. 
 \begin{eqnarray}
 	&&\{B_{1}^{\prime},B_{2}^{\prime}\}\\
  	\nonumber		=&&4\Bigg(\alpha_{1}^2+\frac{\beta_{1}^2}{3}\Bigg)^{2}\{B_{1},B_{2}\}+\frac{4}{3}\beta_{1}^2\Bigg(\alpha_{1}^2+\frac{\beta_{1}^2}{3}\Bigg)\Bigg(B_{1}B_{3}\{B_{2},B_{3}\}+B_{2}B_{3}\{B_{1},B_{3}\}+B_{3}\{B_{1},B_{3}\}B_{2}+B_{3}\{B_{2},B_{3}\}B_{1}\Bigg)\\
 		\nonumber
 		&+&\Bigg(\frac{2}{3}\beta_{1}^2\Bigg)^{2}\Bigg(\{B_{1},B_{2}\}^{3}-3\{B_{1},B_{2}\}-B_{1}B_{3}\{B_{2},B_{3}\}-B_{2}\{B_{1},B_{2}\}B_{2}-B_{1}\{B_{1},B_{2}\}B_{1}-B_{3}\{B_{2},B_{3}\}B_{1}-B_{3}\{B_{1},B_{3}\}B_{2}\\
 		\nonumber
 		&-&B_{2}B_{3}\{B_{1},B_{3}\}+B_{2}\{B_{1},B_{2}\}B_{3}\{B_{2},B_{3}\}+B_{3}\{B_{2},B_{3}\}B_{2}\{B_{1},B_{2}\}+B_{3}\{B_{1},B_{3}\}B_{1}\{B_{1},B_{2}\}+B_{1}\{B_{1},B_{2}\}B_{3}\{B_{1},B_{3}\}+B_{3}\{B_{1},B_{2}\}B_{3}\Bigg)\\
 	\end{eqnarray}
 	The co-efficients $\alpha_{1}>0$, $\beta_{1}\geq 0$. So, the above expression can only be zero when $\{B_{i},B_{j}\}=0$	where $i(j) \in 1,2,3$.
 	
	To derive the optimal quantum value of $(\mathbb{E}^{2})_Q$ in the Eq. (\ref{ebi2}) we use SOS approach where $\omega_{i}^{\prime}=||B_{i}^{\prime}||$ with $i \in \{1,2,3\}$
 where $||.||_{2}$ is the Euclidean norm of a vector.
The optimal quantum value of $(\mathbb{E}^{2})_Q$ can be obtained as   

\begin{eqnarray}
	\label{}
	(\mathbb{E}^{2})_Q^{opt} ={max}\left(\omega_{1}\omega^{\prime}+\omega_{2}\omega_{2}^{\prime}+\omega_{3}\omega_{3}^{\prime}\right)=2(\alpha_{1}^2-\frac{\beta_{1}^2}{3})(\mathbb{E})_Q^{opt}
\end{eqnarray}

\section{Detailed calculation for the sequential quantum value of elegant Bell expression for Alice and Bob$_{3}$}
\label{appC}
	The average reduced state for Alice and Bob$_{3}$ is calculated by using Eq. (\ref{g22}) as
	\begin{eqnarray}
		\label{adp22}
		\nonumber
		\rho_{AB_{3}}&=&\dfrac{1}{3} \sum_{b_{2}\in \{ +,-\}}^{}\sum_{y_{2}=1}^{3} \left(\mathbb{I} \otimes K_{b_{2}|y_{2}}\right)  \rho_{AB_{2}}  \left(\mathbb{I} \otimes K_{b_{2}|y_{2}}\right)\\
		&=& 4\Big(\alpha_{1}^2\alpha_{2}^2+\frac{\beta_{1}^2\beta_{2}^2}{3}\Big) \rho_{AB_{1}} +\frac{4}{3}\Big(\alpha_{1}^2\beta_{2}^2+\beta_{1}^2\alpha_{2}^2\Big)\sum_{{y_{2}}=1}^{3}(\mathbb{I}\otimes B_{y_{2}})\rho_{AB_{1}} (\mathbb{I}\otimes B_{y_2})\\
		\nonumber
		&+&\frac{4\beta_{1}^2\beta_{2}^2}{9}\Big(  B_{1}B_{2}\rho_{AB_{1}}B_{2} B_{1}+B_{2}B_{1}\rho_{AB_{1}}B_{1} B_{2}+B_{1}B_{3}\rho_{AB_{1}}B_{3} B_{1}+B_{3}B_{1}\rho_{AB_{1}}B_{1} B_{3}+B_{2}B_{3}\rho_{AB_{1}}B_{3} B_{2}+B_{3}B_{2}\rho_{AB_{1}}B_{2} B_{3}\Big)
	\end{eqnarray}	
	 
	 	The quantum value of elegant Bell expression between Alice and Bob$_{3}$ from Eq. (\ref{adp22}) is given by
	 
 \begin{eqnarray}
 \nonumber
 	\label{aebi}
 	(\mathbb{E}^{3})_Q
 	&=&max\Bigg(Tr\Big[\rho_{AB_{1}}\Big((A_{1}+A_{2}+A_{3}-A_{4})B_{1}^{\prime\prime}+(A_{1}+A_{2}-A_{3}+A_{4})B_{2}^{\prime\prime}+(A_{1}-A_{2}+A_{3}+A_{4})B_{3}^{\prime\prime}\Big)\Big]\Bigg)
 	\end{eqnarray}
  where
  \begin{eqnarray}
  \nonumber
B_{1}^{\prime\prime}&=&\Big(4\alpha_{1}^2\alpha_{2}^2+\frac{4\beta_{1}^2\beta_{2}^2}{3}+\frac{4\alpha_{1}^2\beta_{2}^2}{3}+\frac{4\beta_{1}^2\alpha_{2}^2}{3}\Big)B_{1}+\Big(\frac{4\alpha_{1}^2\beta_{2}^2}{3}+\frac{4\beta_{1}^2\alpha_{2}^2}{3}+\frac{4\beta_{1}^2\beta_{2}^2}{9}\Big)(B_{2}B_{1}B_{2}+B_{3}B_{1}B_{3})\\
&+&\frac{4\beta_{1}^2\beta_{2}^2}{9}(B_{1}B_{2}B_{1}B_{2}B_{1}+B_{1}B_{3}B_{1}B_{3}B_{1}+B_{3}B_{2}B_{1}B_{2}B_{3}+B_{2}B_{3}B_{1}B_{3}B_{2}),\\
\nonumber
 	B_{2}^{\prime\prime}&=&\Big(4\alpha_{1}^2\alpha_{2}^2+\frac{4\beta_{1}^2\beta_{2}^2}{3}+\frac{4\alpha_{1}^2\beta_{2}^2}{3}+\frac{4\beta_{1}^2\alpha_{2}^2}{3}\Big)B_{2}+\Big(\frac{4\alpha_{1}^2\beta_{2}^2}{3}+\frac{4\beta_{1}^2\alpha_{2}^2}{3}+\frac{4\beta_{1}^2\beta_{2}^2}{9}\Big)(B_{1}B_{2}B_{1}+B_{3}B_{2}B_{3})\\
 	&+&\frac{4\beta_{1}^2\beta_{2}^2}{9}(B_{2}B_{1}B_{2}B_{1}B_{2}+B_{2}B_{3}B_{2}B_{3}B_{2}+B_{3}B_{1}B_{2}B_{1}B_{3}+B_{1}B_{3}B_{2}B_{3}B_{1}),\\
 	\nonumber
 	B_{3}^{\prime\prime}&=&\Big(4\alpha_{1}^2\alpha_{2}^2+\frac{4\beta_{1}^2\beta_{2}^2}{3}+\frac{4\alpha_{1}^2\beta_{2}^2}{3}+\frac{4\beta_{1}^2\alpha_{2}^2}{3}\Big)B_{3}+\Big(\frac{4\alpha_{1}^2\beta_{2}^2}{3}+\frac{4\beta_{1}^2\alpha_{2}^2}{3}+\frac{4\beta_{1}^2\beta_{2}^2}{9}\Big)(B_{1}B_{3}B_{1}+B_{2}B_{3}B_{2})\\
 	&+&\frac{4\beta_{1}^2\beta_{2}^2}{9}(B_{3}B_{1}B_{3}B_{1}B_{3}+B_{3}B_{2}B_{3}B_{2}B_{3}+B_{2}B_{1}B_{3}B_{1}B_{2}+B_{1}B_{2}B_{3}B_{2}B_{1}).
 \end{eqnarray}
 It can also be proved that $\{B_{i}^{\prime\prime},B_{j}^{\prime\prime}\}=0$ with $i,j=1,2,3$ provided $\{B_{i},B_{j}\}=0$.
 \end{widetext}
		
		\end{document}